\documentstyle[12pt,preprint]{aastex}  

\begin{document}

                           % The preamble begins here.
\title{New Extreme Trans-Neptunian Objects: Towards a Super-Earth in the Outer Solar System}
\author{Scott S. Sheppard\altaffilmark{1} and Chadwick Trujillo\altaffilmark{2}}

\altaffiltext{1}{Department of Terrestrial Magnetism, Carnegie Institution for Science, 5241 Broad Branch Rd. NW, Washington, DC 20015, USA, ssheppard@carnegiescience.edu}
\altaffiltext{2}{Gemini Observatory, 670 North A`ohoku Place, Hilo, HI 96720, USA}

\begin{abstract}  % Produces abstract

We are conducting a wide and deep survey for extreme distant solar
system objects.  Our goal is to understand the high perihelion objects
Sedna and 2012 VP113 and determine if an unknown massive planet exists
in the outer solar system.  The discovery of new extreme objects from
our survey of some 1080 square degrees of sky to over 24th magnitude
in the $r$-band are reported.  Two of the new objects, 2014 SR349 and
2013 FT28, are extreme detached trans-Neptunian objects, which have
semi-major axes greater than 150 AU and perihelia well beyond Neptune
($q>40$ AU).  Both new objects have orbits with arguments of perihelia
within the range of the clustering of this angle seen in the other
known extreme objects.  One of these objects, 2014 SR349, has a
longitude of perihelion similar to the other extreme objects, but 2013
FT28 is about 180 degrees away or anti-aligned in its longitude of
perihelion. We also discovered the first outer Oort cloud object with
a perihelion beyond Neptune, 2014 FE72.  We discuss these and other
interesting objects discovered in our ongoing survey.  All the high
semi-major axis ($a>150$ AU) and high perihelion ($q>35$ AU) bodies
follow the previously identified argument of perihelion clustering as
first reported and explained as being from an unknown massive planet
by Trujillo and Sheppard (2014), which some have called Planet X or
Planet 9.  With the discovery of 2013 FT28 on the opposite side of the
sky, we now report that the argument of perihelion is significantly
correlated with the longitude of perihelion and orbit pole angles for
extreme objects and find there are two distinct extreme clusterings
anti-aligned with each other.  This previously unnoticed correlation
is further evidence of an unknown massive planet on a distant
eccentric inclined orbit, as extreme eccentric objects with perihelia
on opposite sides of the sky (180 degree longitude of perihelion
differences) would approach the inclined planet at opposite points in
their orbits, thus making the extreme objects prefer to stay away from
opposite ecliptic latitudes to avoid the planet (i.e. opposite
argument of perihelia or orbit pole angles).

\end{abstract}


\section{Introduction}

\begin{figure}
\epsscale{0.4}
\centerline{\includegraphics[angle=90,totalheight=0.6\textheight]{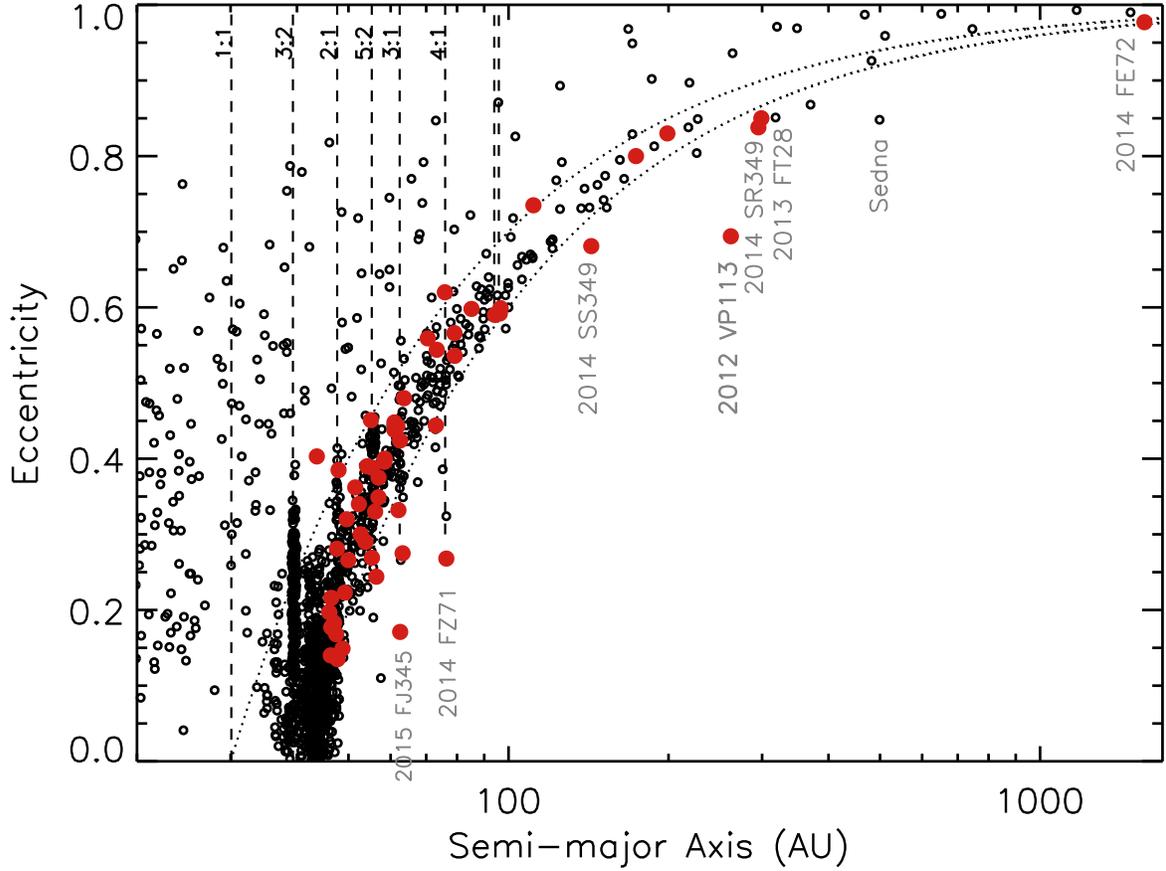}}
\caption{The semi-major axis versus eccentricity for all objects with
  well determined orbits as of July 2016 from the Minor Planet Center.
  Red circles show the new objects discovered in this survey with well
  determined orbits.  Dashed lines show strong mean motion resonances
  with Neptune and dotted lines show constant perihelia of 30 and 40
  AU.  All objects uncertainties are generally smaller than the
  symbols.}
\label{fig:kboea2016} 
\end{figure}

\begin{figure}
\epsscale{0.4}
\centerline{\includegraphics[angle=90,totalheight=0.6\textheight]{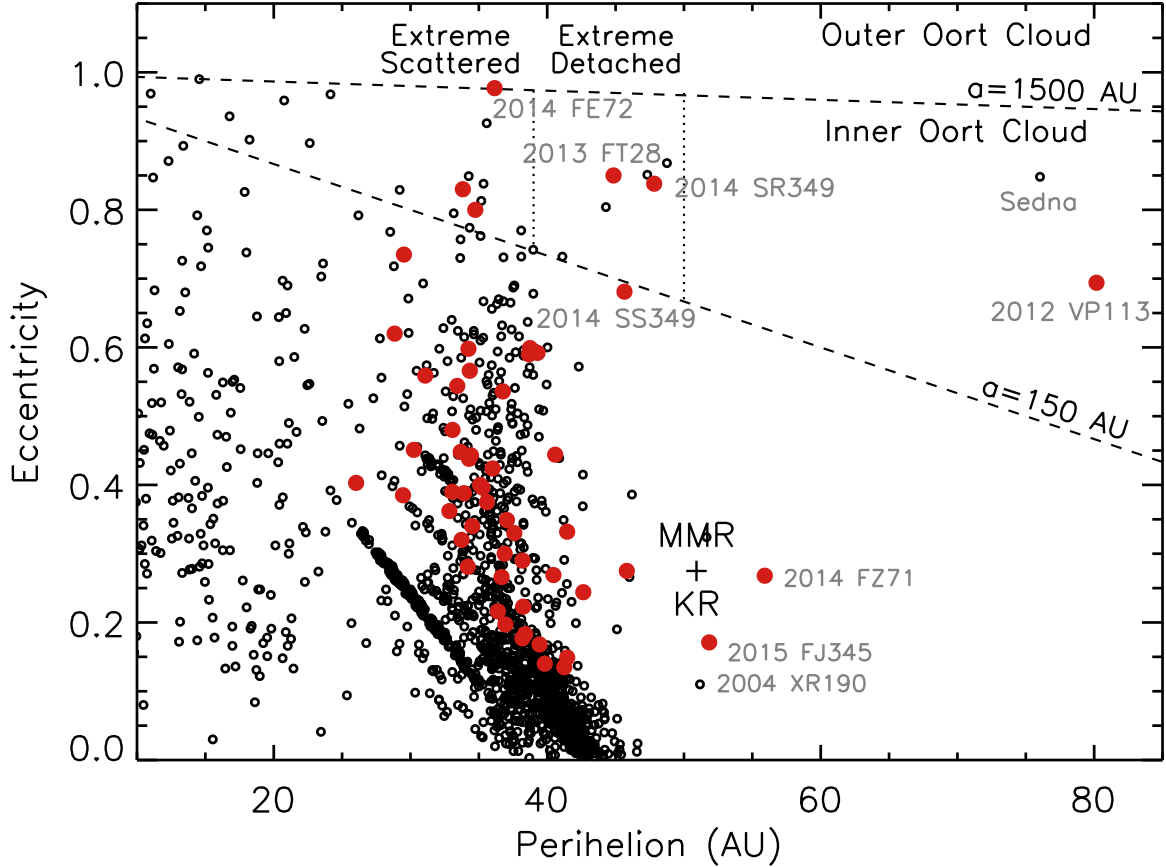}}
\caption{The perihelion versus eccentricity of small outer solar
  system objects with well known orbits as of July 2016 from the Minor
  Planet Center.  Red filled circles are objects discovered during
  this survey with well determined orbits.  Objects above the 150 AU
  semi-major axis dashed line are considered extreme.  Bonafide Inner
  Oort Cloud objects are considered to have perihelion above 50 AU.
  Extreme detached objects are mostly decoupled from the giant planets
  and have perihelion between about 40 and 50 AU and may have a
  similar origin as the IOC objects. Extreme scattered disk objects
  have perihelia below 40 AU and can have significant interactions
  with Neptune.  Outer Oort cloud objects have semi-major axes above
  1500 AU and can have significant interactions with outside forces
  such as the galactic and stellar tides.  Objects with relatively
  high perihelion beyond the Kuiper Belt edge at 50 AU but only
  moderate eccentricity are likely created by a combination of past
  Neptune Mean Motion Resonances (MMR) and the Kozai Resonance (KR)
  and are detailed in Sheppard et al. (2016).  All objects
  uncertainties are generally smaller than the symbols.}
\label{fig:kboeq2016} 
\end{figure}

Extreme trans-Neptunian objects (ETNOs) have orbits that are not
easily explained through the currently observed solar system
configuration.  Objects like Sedna and 2012 VP113 are the two most
prominent members of the ETNOs (Trujillo and Sheppard 2014).  Their
large perihelia of 76 and 80 AU respectively, mean they are decoupled
from the giant planet region, but yet have highly eccentric orbits
(Figures~\ref{fig:kboea2016} and ~\ref{fig:kboeq2016}).  These objects
must have interacted with something in the past to obtain their
distant, eccentric orbits.  Unlike objects in the outer Oort cloud
($a>1500$ AU), which can have their orbits modified by interactions
with the galactic tide, Sedna and 2012 VP113 are too tightly bound to
the Sun to be strongly perturbed by outside forces.  For this reason,
Sedna and 2012 VP113 have been called inner Oort cloud (IOC) objects.

Several viable theories have been put forth to explain the orbits of
IOC objects (Kenyon and Bromley 2004; Morbidelli and Levison 2004;
Brown et al. 2004; Melita et al. 2005; Levison et al. 2010; Kaib et
al. 2011; Jilkova et al. 2015; Portegies Zwart and Jilkova 2015).  The
most popular scenario, until recently, is that the IOC objects were
put into place while the Sun was still within its birth cluster
(Brasser et al. 2012).  During this time, many other young forming
stars were near the Sun and thus outside forces, such as the stellar
tide from nearby stars, were much stronger than today.  The stellar
tide from the Sun's birth cluster could have perturbed objects much
closer to the Sun, possibly down to hundreds of AU if the Sun formed
in a very dense stellar environment (Brasser et al. 2006).  In this
way, IOC objects would be a fossilized population of objects with
major changes to their orbits ending when our Sun left the star
formation cluster.  This would suggest studying the IOC objects would
allow us to gain insights into our Sun's formation environment
(Schwamb et al. 2010).

The IOC objects could also have been created by rogue planets that
were scattered or ejected from the solar system (Morbidelli and
Levison 2004; Brown et al. 2004; Gladman and Chan 2006).  The
involvement of planetary perturbers in the IOC formation gained
additional traction with the observation that Sedna and 2012 VP113,
along with ten other extreme trans-Neptunian objects ($q>35$ AU and
$a>150$ AU), have similarities in their orbits (Trujillo and Sheppard
2014).  These twelve extreme objects cluster within a few tens of
degrees of 340 degrees in their argument of perihelia and are near the
same location on the sky.  Trujillo and Sheppard (2014) demonstrated
this unexpected orbital clustering could be created by an unseen
planet some 2 to 15 Earth masses beyond a few hundred AU, which is
currently shepherding these distant objects into similar orbits likely
through some sort of resonance dynamics.

With the idea that there is an unseen massive planet in the few
hundred AU range from Trujillo and Sheppard (2014), Bromley and Kenyon
(2014) and Kenyon and Bromley (2015) showed how a massive planet might
have obtained such a distant orbit after being scattered out of the
giant planet region and interacting with leftover gas from the planet
formation era.  It also seems plausible that the stellar tide
formation originally envisioned for Sedna would also work for the
formation of the unseen planet's distant orbit.  Thus the unseen
planet may have begun forming near the known giant planets but was
placed into a distant orbit before it accreted a larger amount of
mass.  Because of the quite different semi-major axes of the ETNOs, de
la Fuente Marcos and de la Fuente Marcos (2014a) suggested two planets
instead of just one might be responsible for the clustering of ETNOs.
Several authors further showed how this distant massive planet might
influence other bodies in the solar system such as the extreme
Centaurs (de la Fuente Marcos and de la Fuente Marcos 2014b; Gomes et
al. 2015).  Hees et al. (2014) and Iorio (2014) tried to put limits on
the size and distance of the possible unseen massive planet through
looking at the Cassini spacecraft tracking data and planetary
ephemerides for unexplained gravitational perturbations.

The work by Trujillo and Sheppard (2014) (From here on called TS2014)
identified the surprising result that the ETNOs clustered in their
orbital parameters.  They demonstrated that this clustering was not
from observational biases as most of the ETNOs were discovered at high
ecliptic latitudes.  Further, it was shown the clustering of ETNOs
would only be seen if the mechanism for the clustering of the ETNOs is
still operating, as any orbital clustering from past situations would
disperse over time if no continuing shepherding mechanism is in place.
TS2014 then showed the basic proof of concept that a yet unobserved
outer massive planet beyond a few hundred astronomical units was
likely responsible and could create the ETNO clustering.  TS2014 did
some basic numerical simulations to show that a planet would need to
be larger than about 2 Earth masses to shepherd the orbital angles of
these distant extreme trans-Neptunian objects and that this planet
would need to be beyond about 200 AU to have gone undetected to date.
Although TS2014 presented a 5 Earth mass planet at 250 AU in their
simulation figure, TS2014 specifically mentioned that there were many
possibilities for the orbit of the unknown massive planet and never
specified a best or unique orbit as much more parameter space needed
to be explored then in their initial simulations.

The main parameters identified by TS2014 for the distant outer planet
were a mass between 2 and 15 Earth masses with a distance beyond 200
AU, with a highly inclined 1500 AU orbit being possible.  TS2014
suggested that a more comprehensive, wider orbital parameter analysis
would put tighter constraints on the possible orbits for a unknown
outer planet.  Two years later, Batygin and Brown (2016) (hereafter
BB2016) took this next step with a more rigorous analytic treatment
coupled with dynamical simulations inspired by the ETNO clustering
suggested by TS2014.  BB2016 confirmed that there is indeed a
clustering of ETNOs in argument of perihelion and longitude of
perihelion. They placed stronger constraints on the possible orbit of
the massive unknown planet, showing that the pertuber is likely on an
eccentric orbit that is inclined and anti-aligned with the ETNOs with
a perihelion no closer than about 200 AU and aphelion no further than
about 1500 AU.  The BB2016 result further predicts the highly inclined
and retrograde TNO population (see Gladman et al. 2009) is also
created by the unseen planet.  Malhotra et al. (2016) further looked
into the clustering of the ETNO population and suggest that some of
the ETNOs have orbital period ratios indicative of being in mean
motion resonances with an unknown distant planet.

Some authors have called this possible massive planet beyond a few
hundred astronomical units Planet X or Planet 9.  Fienga et al. (2016)
and Holman \& Payne (2016a,b) have looked further into the Cassini
tracking and planetary ephemerides to put limits on the size and
distance of an unseen massive planet in the outer solar system.
Sheppard et al. (2016) showed that this distant planet likely has no
significant effect on the orbits of objects with semi-major axes
within 100 AU and moderate or lower eccentricities, including objects
in the distant 6:1 Neptune mean motion resonance near 99 AU.
Scattered objects with very distant orbits will likely start to show
effects from the distant planet (Lawler et al. 2016).  Several
additional papers have discussed the possible formation, effects and
composition of the unseen planet in the distant solar system (Kenyon
and Bromley 2016; Fortney et al. 2016; Sivaram et al. 2016; de la
Fuente Marcos and de la Fuente Marcos 2016; Li and Adams 2016; Cowan
et al. 2016; Beust 2016; Linder and Mordasini 2016; Pauco and Klacka
2016; Mustill et al. 2016; Brown and Firth 2016; Whitmire 2016; Gomes
et al. 2016).

We note Madigan and McCourt (2016) recently showed that eccentric
small objects within a massive disk would increase their inclinations
and could clump into similar arguments of perihelia from a
gravitational instability between the bodies and the disk.  Though
intriguing, it is not clear these objects would continue to cluster
once the massive disk is eroded away as quadrupolar interactions with
the inner giant planets would randomize the arguments of perihelia
once again as discussed in TS2014.  Future work is needed, especially
to see if it might be possible that the mass from the more distant
outer Oort cloud could have some dynamical effect on the Inner Oort
Cloud objects.

IOC objects and ETNOs have the potential to inform us about our solar
system's origins and evolution, but yet the number of ETNOs is
currently small.  For this reason, we have been performing the
deepest, widest survey to date for objects beyond the Kuiper Belt edge
at 50 AU.  Here we report several more interesting objects found in
the distant solar system and discuss what new information they might
be telling us about the outer solar system.  This is an ongoing
survey, so this paper highlights the first years of coverage and the
interesting discoveries that have been made.  The discovery and
implications of the inner Oort cloud object 2012 VP113 was reported in
TS2014 and so is only discussed tangentially in
this paper.  In addition, several high perihelion objects ($q>50$ AU)
with only moderate semi-major axes ($50<a<100$ AU) and eccentricities
($e\lesssim 0.3$) were found, such as 2014 FZ71 and 2015 FJ345, and
reported in Sheppard et al. (2016).  These high perihelion but
moderate orbit objects are all near strong Neptune Mean Motion
Resonances with moderate to high inclinations.  These objects likely
obtained their unusual high perihelion orbits through interactions
with the Neptune mean motion resonances and the Kozai resonance and
thus likely have a different history than the extreme objects reported
in this paper.

If there is a distant massive planet shepherding the extreme
trans-Neptunian objects, it implies that the IOC population is not in
fact primordial as their orbits would be modified by the unseen planet
and not static and related to the formation of the Oort cloud as
originally thought. Thus, a different nomenclature than IOC, such as
sednoids, would be best to use when discussing bodies like Sedna and
2012 VP113. Since the possible unseen planet has not been discovered
and no such nomenclature has been agreed upon, in this work we will
use IOC to refer to bodies with extremely high perihelia ($q>70$ AU)
for consistency with past works.

\section{Observations}

The majority of the area surveyed was with the Cerro Tololo
Inter-American Observatory (CTIO) 4 meter Blanco telescope in Chile
with the 2.7 square degree Dark Energy Camera (DECam).  DECam has 62
$2048\times 4096$ pixel CCD chips from Lawrence Livermore Berkeley
Labs with a scale of 0.26 arcseconds per pixel (Flaugher et al. 2015).
The r-band filter was used during the early observing runs (November
and December 2012 and March, May and November 2013) and the ultra-wide
$VR$ filter was used in the later observations (March and September
2014 and April 2015).

\begin{figure}
\epsscale{0.4}
\centerline{\includegraphics[angle=0,totalheight=0.6\textheight]{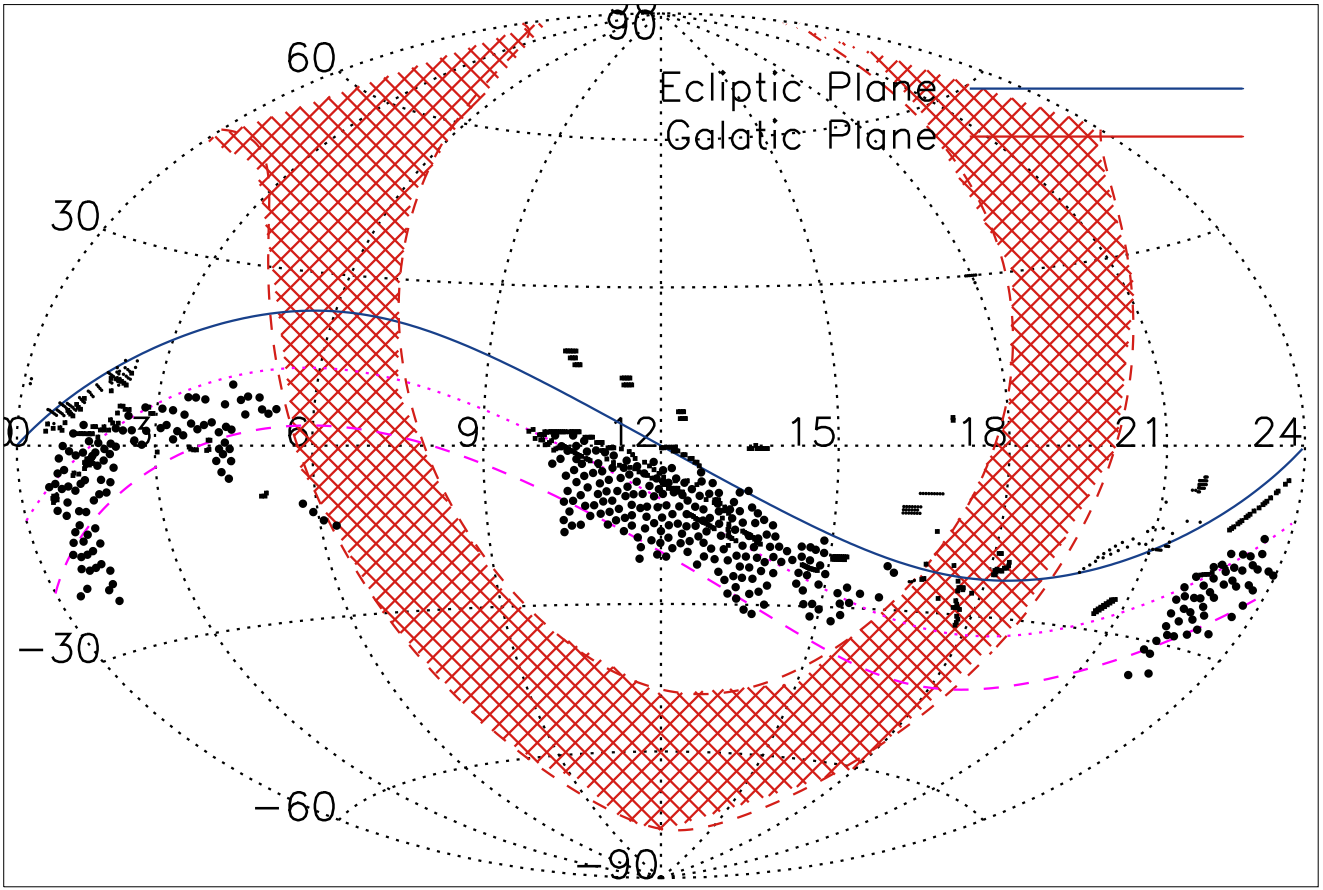}}
\caption{The survey fields obtained using the Dark Energy Camera on
  the CTIO 4m (large circles), MOSAIC2 on the KPNO 4m (large squares),
  IMACS on Magellan (small circles) and SuprimeCam on Subaru (small
  squares) as detailed in Table 1.  The dotted and dashed purple lines
  are respectively 10 and 20 degrees South of the ecliptic.  The
  galactic plane is shown in red to $\pm15$ degrees of the center.  In
  total, we covered about 1080 square degrees during the survey with
  an average of 13 square degrees from the ecliptic.}
\label{fig:MapDECAM} 
\end{figure}

Before DECam became operational the initial inner Oort cloud survey
was begun using the 0.255 square degree SuprimeCam on the 8 meter
Subaru telescope (Miyazaki et al. 2002), the 0.16 sq. deg. IMACS on
the 6.5 meter Magellan telescope (Dressler et al. 2011) and the 0.36
sq. deg. Mosaic-1.1 on the Kitt Peak National Observatory (KPNO) 4
meter Mayall telescope (Jacoby et al. 1998; Wolfe et al. 1998; Sawyer
et al. 2010).  The observing nights and conditions of the survey
fields are shown in Table 1 and Figure~\ref{fig:MapDECAM}.

Usable survey data required no significant extinction from clouds and
seeing less than 1.5 arcseconds at the CTIO 4m and KPNO 4m.  In
general the exposure times were set to reach 24th magnitude with the
r-band filter and 24.5 magnitude with the VR filter during the night.
In the best seeing of 0.8 arcseconds, integration times were around
330 seconds, while in the worst seeing exposure times were up to 700
seconds.  This allowed our survey to obtain a similar depth regardless
of the seeing conditions.  The Subaru and Magellan observations went
deeper, with the target depth of around 25.5 magnitudes in the r-band
and useful seeing being less than 1.0 arcseconds.

The images were all processed in a similar manner.  Images were first
bias-subtracted and then flat fielded with a nightly combination of
twilight flats and master flats created from most of the data in a
single night.  The telescope was autoguided sidereally on field stars
during exposures.  Most fields had three images of similar depth
obtained over a 3 to 5 hour time base (Table 1).  Because of observing
condition changes and night length constraints, a few images had a
slightly smaller time base while others had up to 48 hours between the
first and last image.  Observations were obtained within about 1.5
hours of opposition, which means the dominant apparent motion of any
distant object would be parallactic, and thus inversely related to
distance.

We used a moving object algorithm to detect objects moving at rates
indicative of being beyond Uranus (less than 5 arcseconds per hour).
We were in general sensitive to objects that were about 3 sigma above
the background and moved at least 3 to 4 pixels between the first and
last image of a 3 image set.  The program has been extensively used in
discoveries of distant solar system objects (Trujillo et al. 2001;
Sheppard and Trujillo 2006,2009,2010a) and was slightly modified from
previous versions, such as the pixel scale, field orientations and
saturation limits, to operate on the different wide-field optical
instruments used in this survey (see Trujillo and Sheppard 2014).  The
program first determined all possible objects moving at the desired
rates.  One of the authors then visually inspected all images with the
moving objects reported by the program circled to determine which
objects were real and which were noise from things such as cosmic
rays, saturated stars or bad pixels.  As the eye is very good at
picking out real objects from noise, it was straight forward to
determine which objects to obtain additional observations.  Any
discovery beyond 50 AU was flagged for future recovery observations to
determine if it had an extreme orbit.

Through implanting false objects into our fields, we found most fields
($\sim 90\%$) we were sensitive to objects moving faster than about
0.3 arcseconds per hour, which corresponds to about 500 AU at
opposition.  About $25\%$ of the sky covered in this survey could
detect motions to 700 AU or beyond (generally fields that have over a
4 hour time-base, see Table 1).  Some $14\%$ of the fields had about a
24 hour time-base between the first and last images, allowing objects
out to a few thousand AU to be detected.  False objects of varying
speeds and magnitudes were implanted into a few representative fields
of an observing run and put through the program and visually scanned
just like the rest of the fields to determine the efficiency and
limits of the observing session.

We generally stayed some 15 degrees away from the galactic plane as
shown in Figure~\ref{fig:MapDECAM}.  There are a few fields within the
galactic plane that used known dark clouds to search areas with low
stellar background as described in Sheppard and Trujillo 2010b.  Most
of the survey fields were between 5 and 20 degrees from the ecliptic,
with an average of about 13 degrees from the ecliptic.  The fields
have a fairly uniform longitudinal coverage and are mostly in the
southern sky (Figure~\ref{fig:MapDECAM}).  This work is about the
first 1080 square degrees covered to date to over 24th magnitude
(Subaru$=40$, Magellan$=60$, KPNO$=11$ and DECam$=970$ square
degrees).

\section{Observational Bias Simulator For Population and Orbit Statistics}
\label{PopulationStatistics}

We created an observational bias simulator to place population
estimates and compute orbital statistics on the extreme and the inner
Oort cloud objects.  For our input, we have taken the positions and
depths of fields listed in Table 1. Since mean anomalies were
randomized for all simulated populations, ephemerides were computed
for only a single date to speed execution time. A simulated object was
considered to be detected when located in one of our survey fields
with apparent brightness greater than our survey limit. For detection,
an object also had to have heliocentric distance greater than our
survey lower limit for follow-up, 50 AU, and distance less than 500
AU, where opposition apparent motion falls below our $\sim 0.28$
arcsec per hour detection limit for many of our fields. We have not
included data from other surveys in this work mainly because their
field locations and depths are not known to us and they typically used
smaller observation time-base and smaller telescopes and were therefore
less sensitive to very distant objects.

We had several basic input parameters that were used for all of our
populations. Since very few numbers of IOCs and ETNOs are known, our
size distribution is based on the hot component of the TNOs, which
have a luminosity function slope of $\alpha_h = 0.8$ (Petit et al.
2011; Schwamb et al. 2014). This is the equivalent of a differential
power law to the size distribution of $q' = -5$ (Trujillo 2000). Both
the IOCs and ETNOs could share origins with the hot component of the
TNOs since none of these populations is thought to have formed in
their current locations (Tsiganis et al 2005).  As shown in section
5.4 below, the IOCs and ETNOs appear to have a similar cumulative
luminosity function slope as the other small body populations and thus
likely follow the same $q'\sim -5$ size distribution.  For completeness,
we also simulated a size distribution of $q' = -4$.

As mentioned in section 4 below, we find the average inclination of
the ETNOs to be similar as that found for the scattered disk objects
by Gulbis et al. (2010).  For this reason the inclination distribution
is assumed to be similar to the scattered TNOs with an average around
$\mu = 19.1\deg$ and a gaussian of $\sigma = 6.9\deg$ (Gulbis et
al. 2010). Other simulation-wide parameters are largely assumed since
there is very little observational constraint on them and are detailed
later.

\section{Results: New Extreme Trans-Neptunian Objects}

\begin{figure}
\epsscale{0.4}
\centerline{\includegraphics[angle=90,totalheight=0.6\textheight]{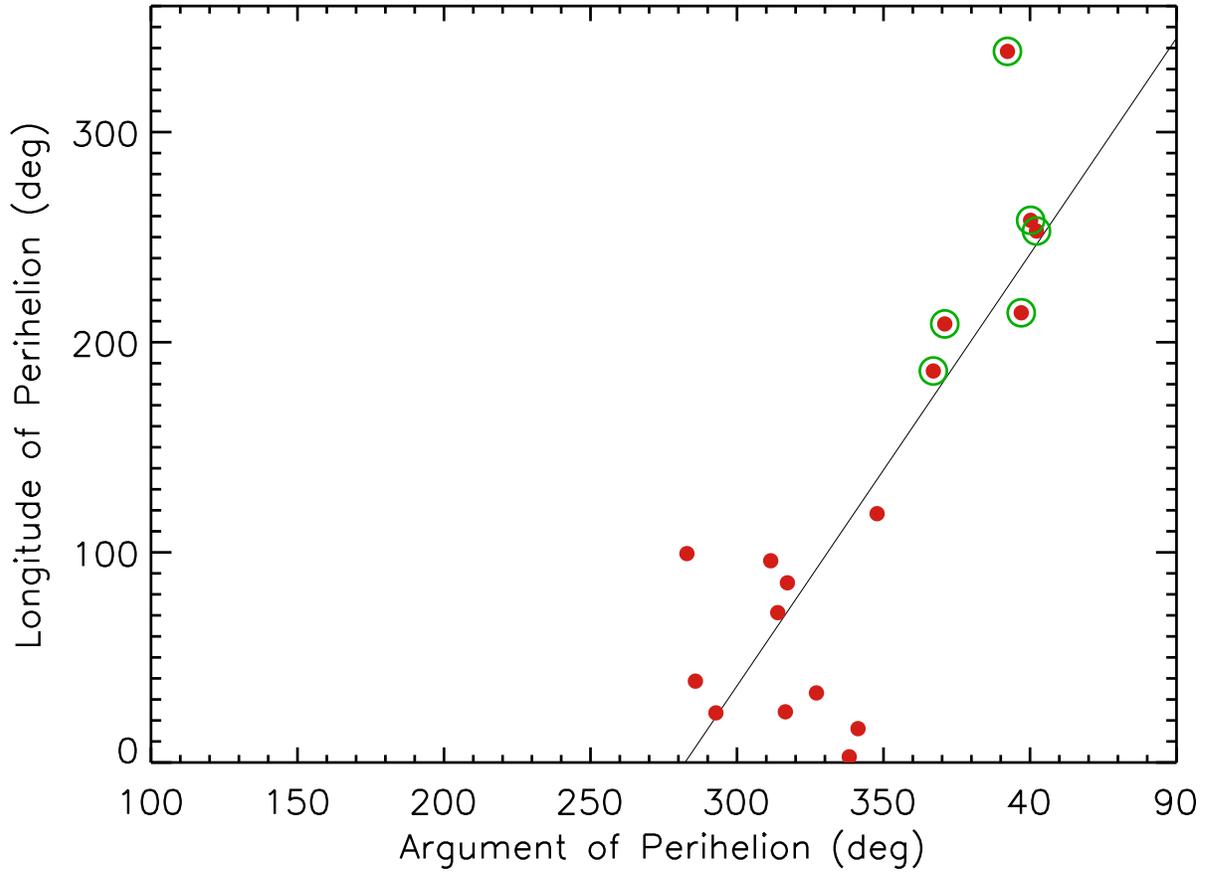}}
\caption{Argument of perihelion versus longitude of perihelion for
  extremely distant trans-Neptunian objects.  There is a clear
  correlation between these two angles at the 99.99\% level.  As
  explained in the text, this is further evidence of a massive
  inclined planet at a few hundred AU. Green circles show the objects
  in the secondary anti-longitude of perihelion cluster.}
\label{fig:distance2016}     
\end{figure}

\begin{figure}
\epsscale{0.4}
\centerline{\includegraphics[angle=90,totalheight=0.6\textheight]{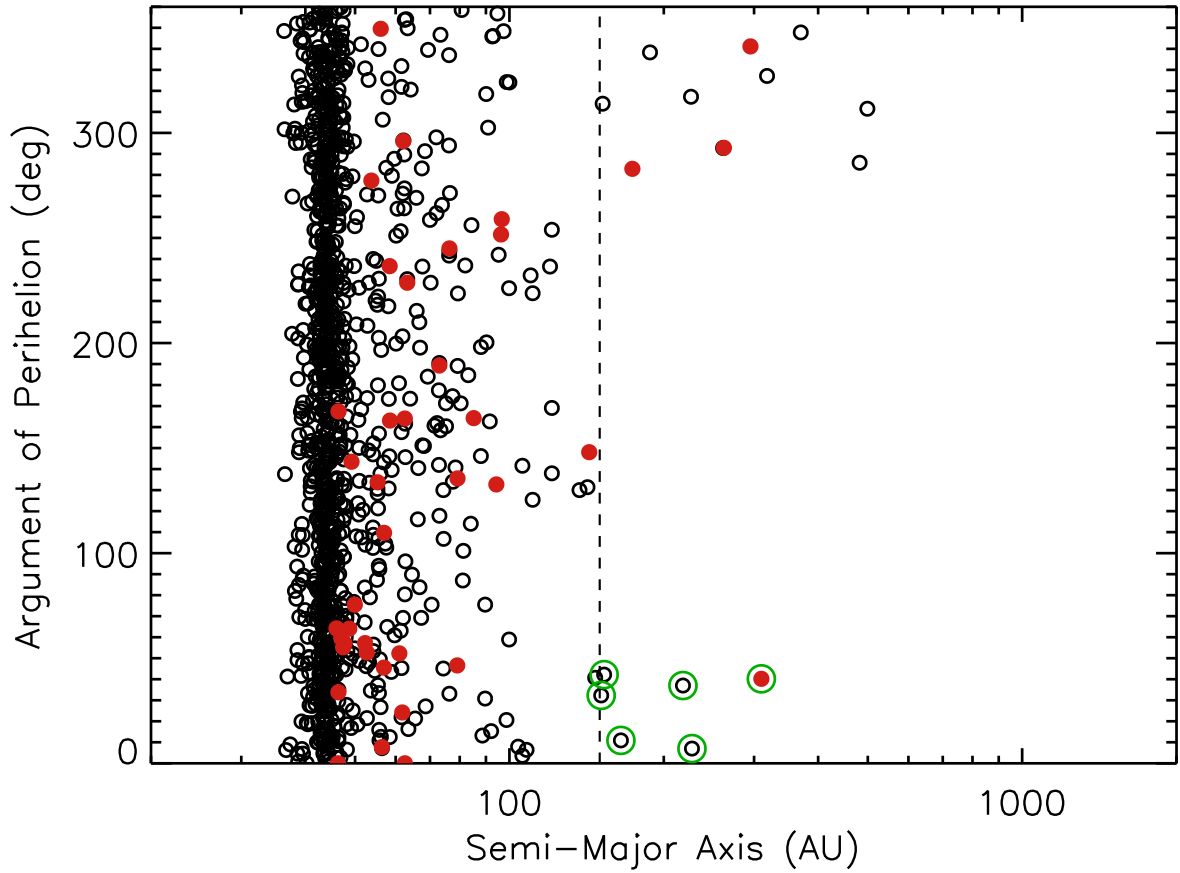}}
\caption{Semi-major axis versus argument of perihelion for all objects
  with perihelia greater than 35 AU.  There is a noticeable clustering
  between 290 and 40 degrees.  Red symbols are new objects discovered
  in this survey.  Green circles show the objects that are in the
  secondary anti-longitude of perihelion clustering group.} %*****
\label{fig:kboaw2016more} 
\end{figure}

\begin{figure}
\epsscale{0.4}
\centerline{\includegraphics[angle=90,totalheight=0.6\textheight]{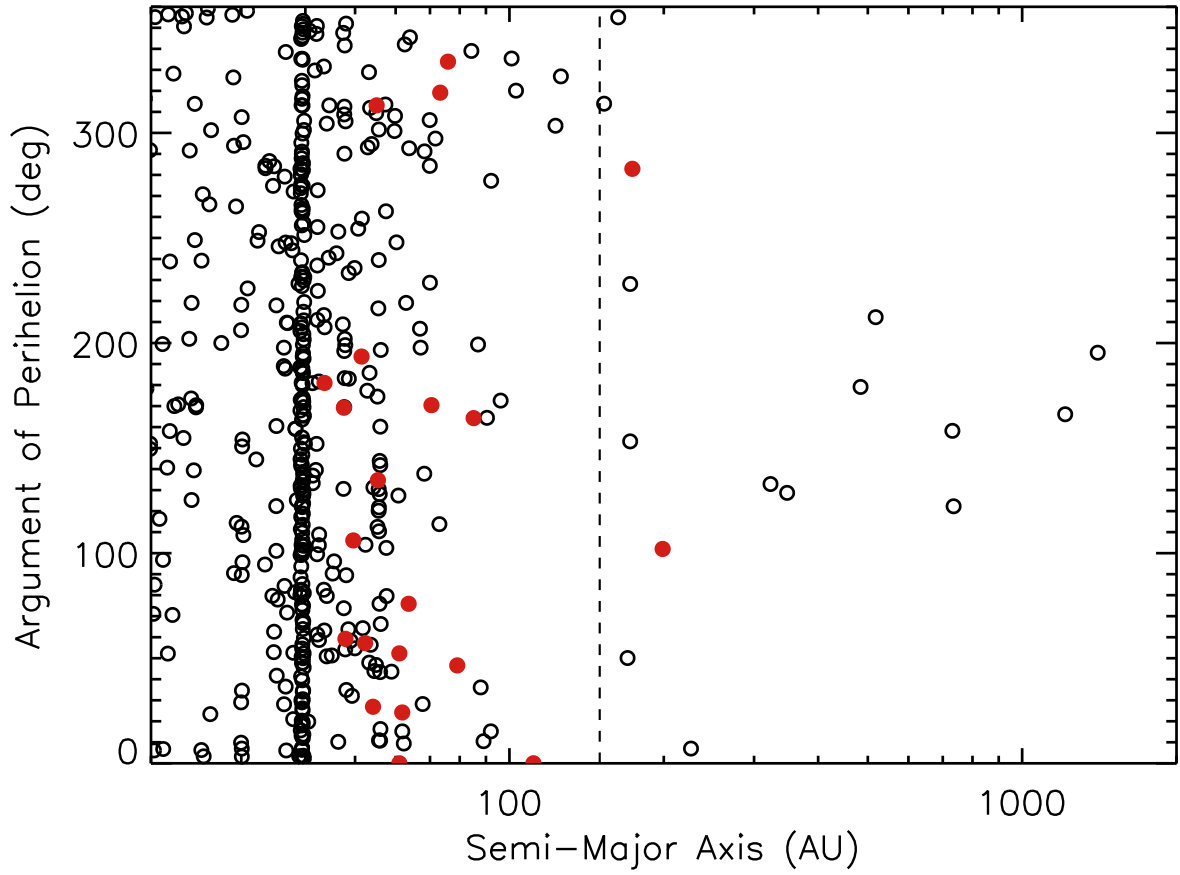}}
\caption{Semi-major axis versus argument of perihelion for all objects
  with perihelia less than 35 AU.  There is a noticeable clustering
  between 100 and 200 degrees, which is the opposite of
  Figure~\ref{fig:kboaw2016more}.}  %*****
\label{fig:kboaw2016less} 
\end{figure}

\begin{figure}
\epsscale{0.4}
\centerline{\includegraphics[angle=90,totalheight=0.6\textheight]{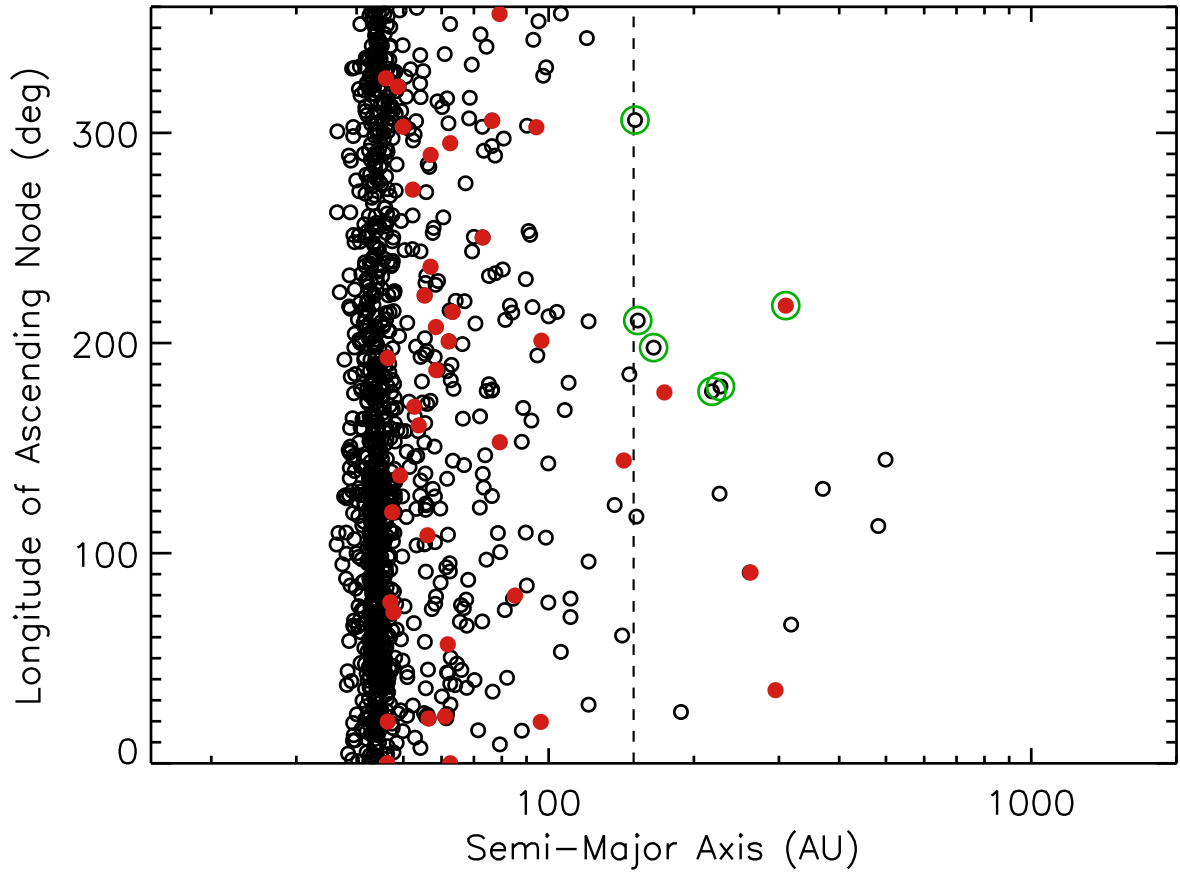}}
\caption{Semi-major axis versus longitude of the ascending node with
  perihelia greater than 35 AU.  Red symbols are new objects
  discovered in this survey.  Green circles show the objects that are
  in the secondary anti-longitude of perihelion clustering group.}
\label{fig:kboaLong2016more}    %*****
\end{figure}

\begin{figure}
\epsscale{0.4}
\centerline{\includegraphics[angle=90,totalheight=0.6\textheight]{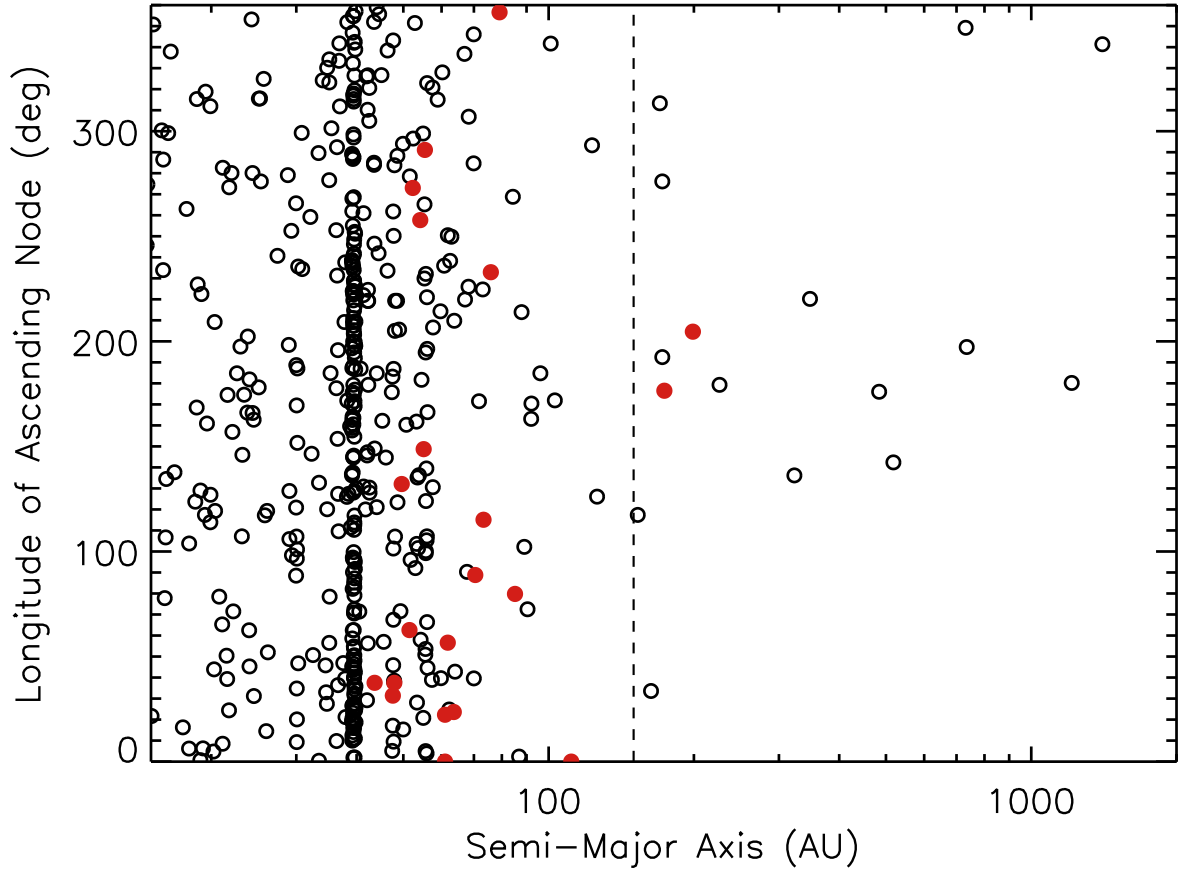}}
\caption{Semi-major axis versus longitude of the ascending node with
  perihelia less than 35 AU.  There is a clustering near 180 degrees.}
\label{fig:kboaLong2016less}   %*****
\end{figure}

\begin{figure}
\epsscale{0.4}
\centerline{\includegraphics[angle=90,totalheight=0.6\textheight]{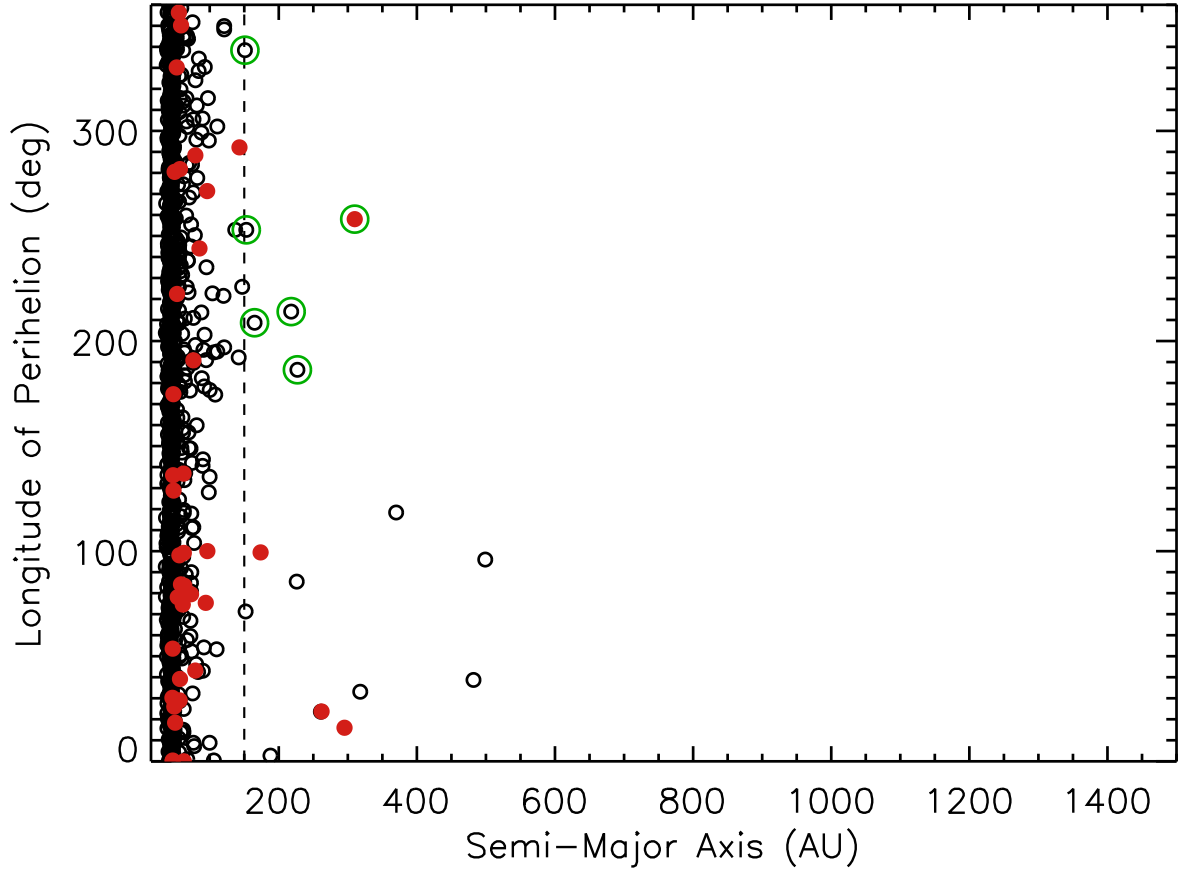}}
\caption{Semi-major axis versus longitude of perihelion with perihelia
  greater than 35 AU.  There is a main clustering between 0 and 130
  degrees for most extreme objects with a possible second clustering
  for lower semi-major axes between 200 and 260 degrees (circled in
  green).  2013 FT28 does not follow the longitude of perihelion
  clustering of the other extreme TNOs with the highest semi-major
  axes, but may fall into the secondary clustering region.  Red dots
  are new discoveries from this survey.}
\label{fig:kboaOmega2016more} 
\end{figure}

\begin{figure}
\epsscale{0.4}
\centerline{\includegraphics[angle=90,totalheight=0.6\textheight]{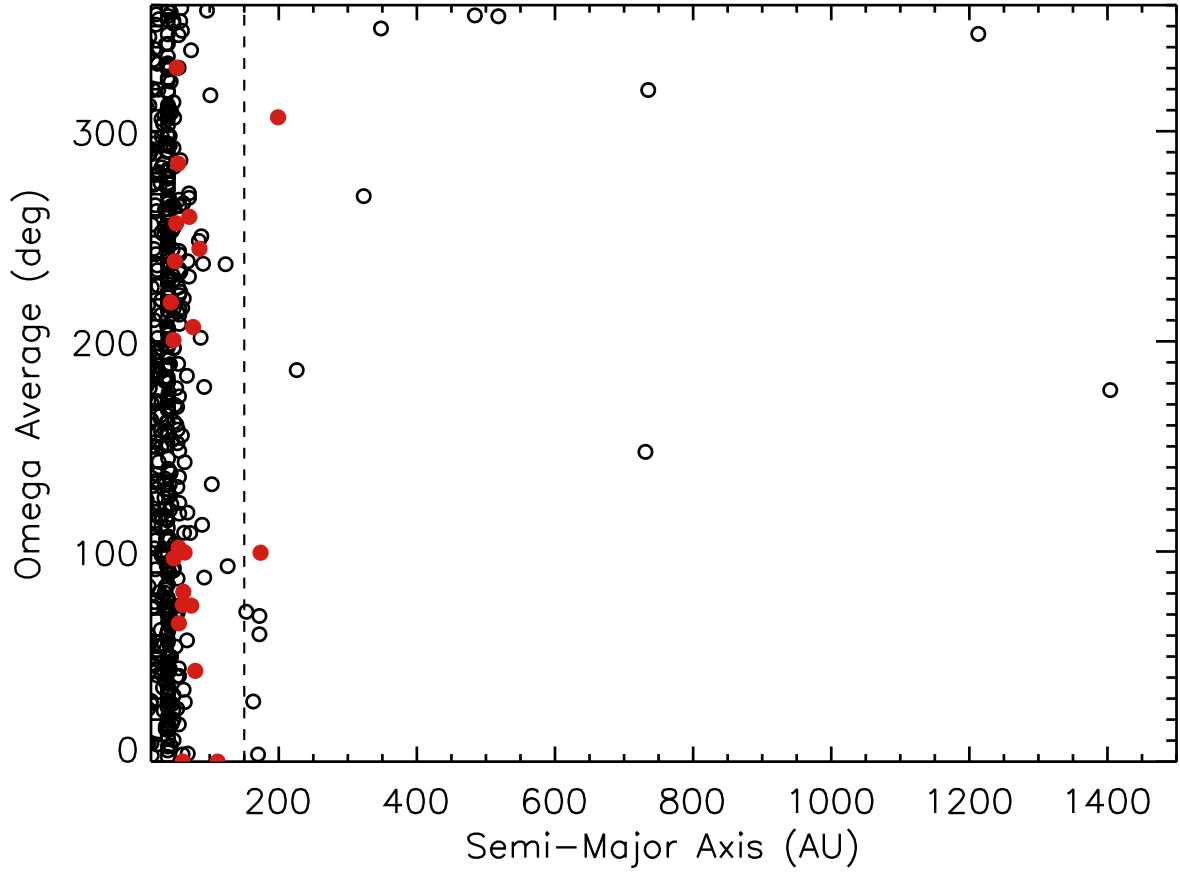}}
\caption{Semi-major axis versus longitude of perihelion with perihelia
  less than 35 AU.  There are marginal clusterings near 0 degrees and
  180 degrees.}
\label{fig:kboaOmega2016less} 
\end{figure}

This survey has now discovered new ETNOs that could have an origin
similar to the IOC objects and thus could constrain observations of
dynamical processes in the outer solar system beyond the Kuiper Belt.
The orbital elements of the newly discovered objects plotted with the
well known outer solar system objects are shown in
Figures~\ref{fig:distance2016} to~\ref{fig:kboaOmega2016less}.  The
figures are divided between objects with perihelia greater than and
less than 35 AU.  This perihelion cutoff value might be better chosen
even great than 35 AU as current Neptune interactions are likely
significant up to at least 38-40 AU (Gomes et al. 2008) or 41 AU
(Brasser and Schwamb 2015) for objects with semi-major axes greater
than 250 AU.  We use 35 AU simply because from an observational
perspective, this is where the orbital data for extreme objects
appears to start to cluster for the argument of perihelion.  We note
that since the extreme objects start to cluster at a perihelion of 35
AU, where Neptune interactions are still fairly strong, this suggests
the mechanism that is creating the clustering is stronger than the
Neptune interactions beyond a perihelion of 35 AU.

The ETNOs other than Sedna and 2012 VP113 have perihelia less than 50
AU and thus are not as decoupled from the giant planet region as the
IOC objects and thus their origins could be similar to those of Sedna
and 2012 VP113 or they could have more significant modifications from
Neptune interactions.  It appears the few objects with perihelia above
40 AU but below 50 AU with high eccentricities are likely to have a
similar origin as Sedna and 2012 VP113 as their interactions with
Neptune should be minimal (Gladman et al. 2002; Gomes et al. 2008;
Chen et al. 2013; Brasser and Schwamb 2015).  In this work, we call
objects with $40<q<50$ AU and semi-major axes greater than 150 and
less than 1500 AU extreme detached trans-Neptunian objects (EDTNOs).
These extreme detached objects, though they have smaller perihelia
than the IOCs, spend the majority of their time between where Neptune
and the known giant planets dominate and where the galactic tide
becomes significant (50 to 1500 AU).  Thus if there is any massive
object still in this gravitationally depleted zone, it should show up
as a signature on the orbits of these EDTNOs (Trujillo and Sheppard
2014).  Their colors seem to be moderately red and thus similar to
scattered disk objects (Sheppard 2010).  New discoveries 2014 SR349 and
2013 FT28 are EDTNOs and have secure orbits as they have been observed
for 2 and 4 oppositions, respectively.  It is interesting to note that
all of the known EDTNOs have perihelia near 45 AU while both known
IOCs have perihelia near 80 AU.  It is low number statistics, so not
much can be said, but their is obvious gaps between the extreme
scattered objects, EDTNOs and IOCs in terms of perihelia
(Figure~\ref{fig:kboeq2016}).

The inclinations of the known EDTNOs and IOCs appear to be moderate
with all but Sedna being between 17 and 26 degrees.  We find the seven
IOCs and EDTNOs have an average inclination of about 20 degrees.  This
is very similar to the inclination average of about 19 degrees found
for the scattered disk objects by Gulbis et al. (2010).

\subsection{Argument of Perihelion Clustering}

Trujillo and Sheppard (2014) found that all ETNOs have similar
arguments of perihelia and predicted a massive planet between some 200
to 1500 AU was shepherding the ETNOs into these similar orbital
angles.  Interestingly, both of the new EDTNO discoveries 2014 SR349
and 2013 FT28 have arguments of perihelia in the same range as the
other three known extreme detached objects (2010 GB174, 2004 VN112 and
2000 CR105) and two inner Oort cloud objects (2012 VP113 and Sedna) as
well as almost all the ETNOs with perihelia greater than Neptune's
(Figure~\ref{fig:kboaw2016more}).  This strengthens the case that the
clustering noticed in TS2014 is real and is statistically significant
around the $6\sigma$ level.

With the discovery of 2013 FT28, we now show there is a significant
correlation between argument of perihelion and longitude of perihelion
angles for ETNOs (Figure 4).  ETNOs with longitude of perihelia
between 0 and 120 deg have arguments of perihelia between 280 and 360
deg.  ETNOs with longitude of perihelia on the other side of the sky
between 180 and 330 deg have arguments of perihelia between 0 and 40
deg.  Using a linear Pearson correlation, we find a Pearson
correlation coefficient of 0.82 using a sample size of 15 ETNOs with
$a>150$ AU and $q>35$ AU, which gives a significance of the
correlation between $\bar{\omega}$ and $\omega$ of around the 99.99\%
level.  These two anti-clustered groupings of ETNOs can be further
seen in their clustering in the Longitude of the Ascending Node angle
(Figure 7) and orbit polar orientation angles, where the $\omega$ 0 to
40 deg group all come to perihelion above the ecliptic and the
$\omega$ 280 to 360 deg group all below the ecliptic.

These correlations are further evidence of an unknown massive planet,
with the distant eccentric inclined orbit shown by BB2016 possible.
This is because extreme eccentric objects with perihelia on opposite
sides of the sky (180 degree longitude of perihelion differences)
would approach the inclined planet at opposite points in their orbits
(see Figure 12).  Thus for the extreme objects to stay as far away as
possible from the planet, they would prefer opposite ecliptic
latitudes of the planet in order to avoid the planet above or below it
(i.e. opposite argument of perihelion or orbit pole angles).  Since
the definition of the pole angle is based on the perihelion and the
two groups have opposite perihelia locations, they would also have
opposite pole angles or arguments of perihelia.  The two ETNO groups
seem to show some sort of resonance behaviour with multiple orbital
angles set to keep them away from the planet.

It is interesting to see that for the lower perihelion objects, $q <
35$ AU, where strong Neptune interactions should be at work, there is
also clustering in the argument of perihelion for high semi-major axes
($a>200$ AU), but opposite or 180 degrees away from the ETNOs
(Figure~\ref{fig:kboaw2016less}).  This possible anti-correlation for
$q<35$ AU objects from the $q>35$ AU objects with $a>150$ AU is
currently unexplained.

\subsubsection{Argument of Perihelion Observational Biases}

There are two biases that could crop up in the argument of perihelion
angle when discovering extreme objects.  As noted in TS2014, surveys
that are preferentially done near the ecliptic will be heavily biased
towards finding objects with arguments of perihelion near 0 and 180
degrees as that is when those objects would be at perihelion and near
the ecliptic and brightest.  This 0 and 180 degree bias goes away as a
survey covers more area away from the ecliptic and becomes minimal
once a survey averages at least 10 degrees from the ecliptic.  As our
survey averages some 13 degrees off the ecliptic, there should be
little to no 0 and 180 degree argument of perihelion bias.  Further,
even with the above 0 and 180 degree bias, one would still be equally
likely to find ETNOs with arguments of perihelion near 180 degrees,
where none or known, as near 0 degrees, where all are known.  TS2014
show the above bias is not a factor in the observed clustering because
objects found at high ecliptic latitudes, like most EDTNOs, will not
have an observational bias in their argument of perihelion value.
Both of the two new extreme detached objects reported here further
refute any bias as both were found over five degrees from the ecliptic
(17 and 7 degrees).  We model observational biases specific to our
survey below and show that the argument of perihelion clustering is
extremely unlikely to be caused by chance.

\begin{figure}
\epsscale{0.4}
\centerline{\includegraphics[angle=0,totalheight=0.60\textheight]{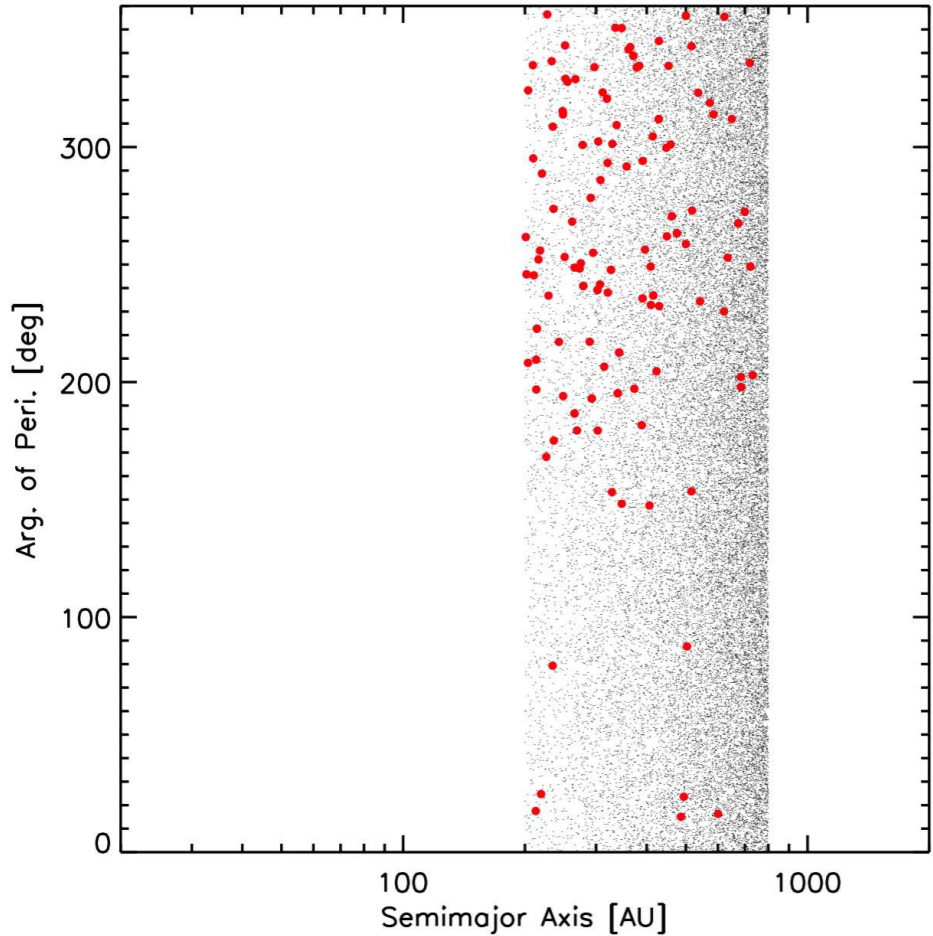}}
\caption{Semi-major axis versus argument of perihelion for a symmetric
  ETNO distribution.  Small black dots are simulated objects and
  larger red dots are objects that would have been detected in our
  survey given our sky coverage.  Because most of our survey was done
  South of the ecliptic, we are less sensitive to detecting objects
  with arguments of perihelion between 0 and 180 degrees.  Despite
  this secondary bias, as explained in the text, we still find strong
  evidence for the argument of perihelion clustering for ETNOs near 0
  degrees where all are known as we are still just as likely to find
  ETNOs near 180 degrees, where none are known.}
\label{fig:omegaa} 
\end{figure}

A secondary argument of perihelion discovery bias exists if the survey
fields are preferentially off the ecliptic in either the North or the
South.  Our survey reported here has mostly been obtained in the
Southern Hemisphere as most observations have been from the CTIO 4m
telescope with DECam that is located at about 30 degrees South
latitude (Figure~\ref{fig:MapDECAM}).  As our fields are mostly South
of the ecliptic we show this bias in Figure~\ref{fig:omegaa}.  That
is, having mostly fields South of the ecliptic we are more sensitive
to ETNOs with arguments of perihelion between 180 and 360 degrees, as
they are more likely to come to perihelion and thus be brightest South
of the ecliptic.  But this secondary bias is not a major concern in
showing that the ETNOs cluster in their arguments of perihelion with a
range between about 290 and 40 degrees.  That is because our survey,
even with this secondary argument of perihelion bias, would be equally
likely to find ETNOs between 90 and 270 degrees, where none are known,
as 270 to 90 degrees, where all are known.  Our ongoing survey is now
also using the Subaru telescope in the North which should help remove
this secondary bias from our observations.  In addition, many of the
ETNOs were discovered in surveys that did cover large Northern sky
areas. We conclude there should be no major biases in the arguments of
perihelion for the known ETNOs and thus they show strong clustering in
their arguments of perihelion between about 290 and 40 degrees.

\subsection{Longitude of Perihelion Clustering}

Trujillo and Sheppard (2014) commented that both Sedna and 2012 VP113
are currently at similar locations on the sky and thus the IOCs may
not be longitudinally symmetric.  TS2014 further mentioned that if
this was found to be true through the discovery of other IOC objects
it would strongly constrain any formation scenario.  This similar
location on the sky seems unlikely to happen by chance and suggests a
clustering in perihelion location.  This was looked at in detail by
Batygin and Brown (2016) to try and constrain what any distant
planet's orbit may look like.  But the similar location on the sky
could be a selection effect as there is a possible bias in longitude
of perihelion observations (Figure 14).  This is because objects are
most likely to be discovered at perihelion when they are brightest.
The bias comes in because most surveys stay away from the galactic
plane of the Milky Way to avoid confusion with background stars
(around 5 to 8 hours and 16 to 19 hours in Right Ascension near the
ecliptic, see Figure~\ref{fig:MapDECAM}).  Thus there are two main
times of the year that surveys for outer solar system objects are most
efficient, which could possibly cause a bias in the longitude of
perihelion discoveries.

\begin{figure}
\epsscale{0.4}
\centerline{\includegraphics[angle=90,totalheight=0.6\textheight]{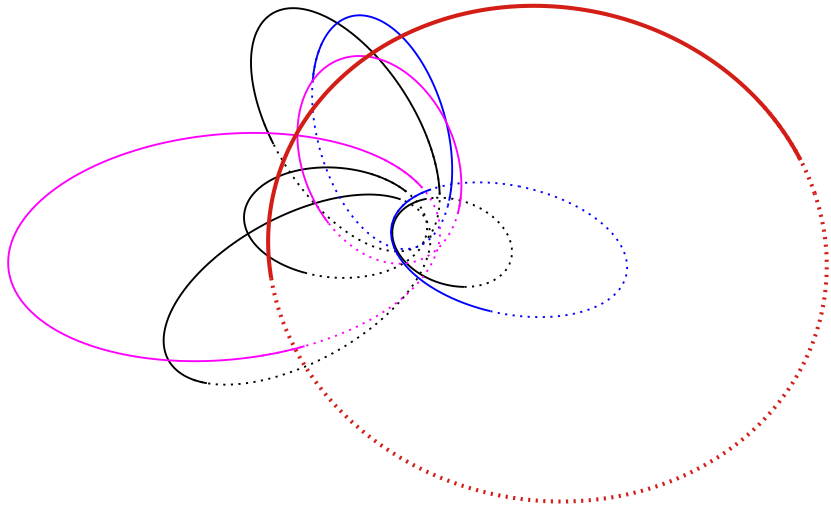}}
\caption{A top down view of the orbits of the extreme detached
  trans-Neptunian objects (black) and Inner Oort Cloud objects
  (purple) with $a>150$ and $q>40$ AU.  The new extreme detached
  objects discovered in this survey are shown in blue (2014 SR349 and
  2013 FT28).  The BB2016 planet orbit prediction is shown in red.
  The dashed part of an orbit is when the object is below the
  ecliptic.  The new object 2013 FT28 does not fit the longitude of
  perihelion trend shown by the other EDTNOs and IOC objects, but its
  location nearly 180 degrees away suggests a secondary anti
  clustering in longitude of perihelion for ETNOs.  We find a distinct
  180 deg anti group amongst the ETNOs that also has opposite argument
  of perihelia and orbit pole angles from the main clustering.  The
  two ETNO groups seem to show some sort of resonance behaviour with
  multiple orbital angles set to keep them away from the
  planet. Though we only plot the stable ETNOs here for clarity, all
  ETNOs from Table 3, stable and unstable, in both clusters show
  similar behaviour befitting the cluster group.}
\label{fig:ETNOplan2016} 
\end{figure}

Although the survey that found Sedna was all sky (Brown et al. 2004),
the initial survey that found 2012 VP113 was not, and was only
conducted in a similar location as Sedna (Trujillo and Sheppard 2014).
This was a coincidence simply because that was the area observable for
the first telescope observations of this survey when 2012 VP113 was
found and thus it was highly biased in the longitude of perihelion
angle of any discovered extreme objects.  Thus the initial discovery
of 2012 VP113 could not be used for the longitude of perihelion
clustering as it was found in a highly biased location of the TS2014
survey.  This is why TS2014 stated it was interesting that Sedna and
2012 VP113 shared similar locations on the sky but further discoveries
would need to be made to look at this point further.  Now with these
new results presented here, the longitude of perihelion bias can be
removed from 2012 VP113 as this survey has now covered similar amounts
of sky on the other side of the galactic plane from Sedna's location
and discovered new EDTNOs.  Not including 2012 VP113, one extreme
detached object was found on each side of the galactic plane during
this survey.  2014 SR349, which has the fourth highest perihelion of
any ETNO, including Sedna and 2012 VP113, was found with a similar
argument of perihelion and longitude of perihelion as Sedna and 2012
VP113.  2013 FT28 was found on the other side of the sky with a
longitude of perihelion some 180 degrees away from Sedna and 2012
VP113 (Figure~\ref{fig:ETNOplan2016}).

It is interesting that 2013 FT28's longitude of perihelion is nearly
180 degrees away from the general center of the other longitude of
perihelion extreme objects.  BB2016 do suggest that if there is a
giant planet on an eccentric orbit creating the longitude of
perihelion clustering for extreme KBOs, there should also be objects
180 degrees away in longitude of perihelion.  BB2016 predict such
objects would be on much lower eccentricity orbits with high
semi-major axes and thus much higher perihelia.  In this regard, 2013
FT28 does not agree with the specific predictions made by BB2016.  In
a second paper, Brown and Batygin (2016) show a possible secondary
longitude of perihelion clustering 180 degrees away from the main
longitude of perihelion clustering for some simulations.  This smaller
population secondary longitude of perihelion clustering should
generally take place at lower semi-major axes than most of the main
clustering and it appears 2013 FT28 is in this secondary longitude of
perihelion clustering.

\subsubsection{Stability of 2013 FT28}
There must be some objects that happen to be passing through the ETNOs
region temporarily and are not on stable orbits, especially if there
is an unknown massive planet beyond a few hundred AU scattering
objects.  The number of these passerbys is unknown.  We looked deeper
into 2013 FT28's orbit with some numerical simulations using the Mercury
integrator (Chambers 1999).  We cloned the orbit of 2013 FT28 a hundred
times to cover the one sigma parameter space of its orbital
uncertainty, which is low with about 3 years (2013-2016) of
observational arc.  Our simulations used the four giant planets and
added the mass of the terrestrial planets to the Sun.  The time step
was 20 days and all integrations ran for 4 billion years.

We noticed that 2013 FT28's orbit is mostly stable but shows some
movement in its semi-major axis over time, suggesting some interactions
with the known giant planets, even though it has a high perihelion and
large semi-major axis.  In our numerical simulations about 30 percent
of the time we found the semi-major axis of 2013 FT28 moved over 50 AU
over the age of the solar system.  Though none of the clones were lost
from the solar system, significant semi-major axis movement suggests
significant interactions with the known giant planets.  In addition,
the semi-major axis of the 2013 FT28 clones that didn't have moderate
semi-major axis movement still showed some jumpiness in their
semi-major axes suggesting some sort of resonant interactions with the
giant planets (Figure~\ref{fig:orbitsall2016}).

This semi-major axis movement is unlike the 2014 SR349 numerical
simulations that found all possible orbits to be stable in semi-major
axis within a few AU over the age of the solar system
(Figure~\ref{fig:orbitsall2016}).  It is thus possible that 2013 FT28
has some sort of resonance interaction with the currently known giant
planets, making its orbit less stable.  The semi-major axis of the
orbit of 2013 FT28 is known to within a few AU, but future
observations might give an orbit that is more in the stable regime.
We note that 2013 FT28 has a perihelion that is well beyond the 40-41
AU limit usually imposed for strong Neptune interactions for objects
with high semi-major axes (Gomes et al. 2008; Brasser and Schwamb
2015).

\subsection{Stable and Unstable ETNOs}
BB2016 used only six of the twelve extreme objects identified in
TS2014 to look for further evidence and possible orbits of this
putative distant massive planet.  BB2016 state in their section 2 that
only stable objects over 4 billion years are useful in their analysis
and is the reason they use only six of the 12 extreme objects. Stable
to them was any extreme object that did not move significantly in its
semi-major axis for 4 billion years.  This limited BB2016 to the six
extreme objects, which according to their Figure 1, were 2012 VP113,
Sedna, 2010 GB174, 2004 VN112, 2000 CR105 and 2010 VZ98.

\begin{figure}
\epsscale{0.4}
\centerline{\includegraphics[angle=0,totalheight=0.55\textheight]{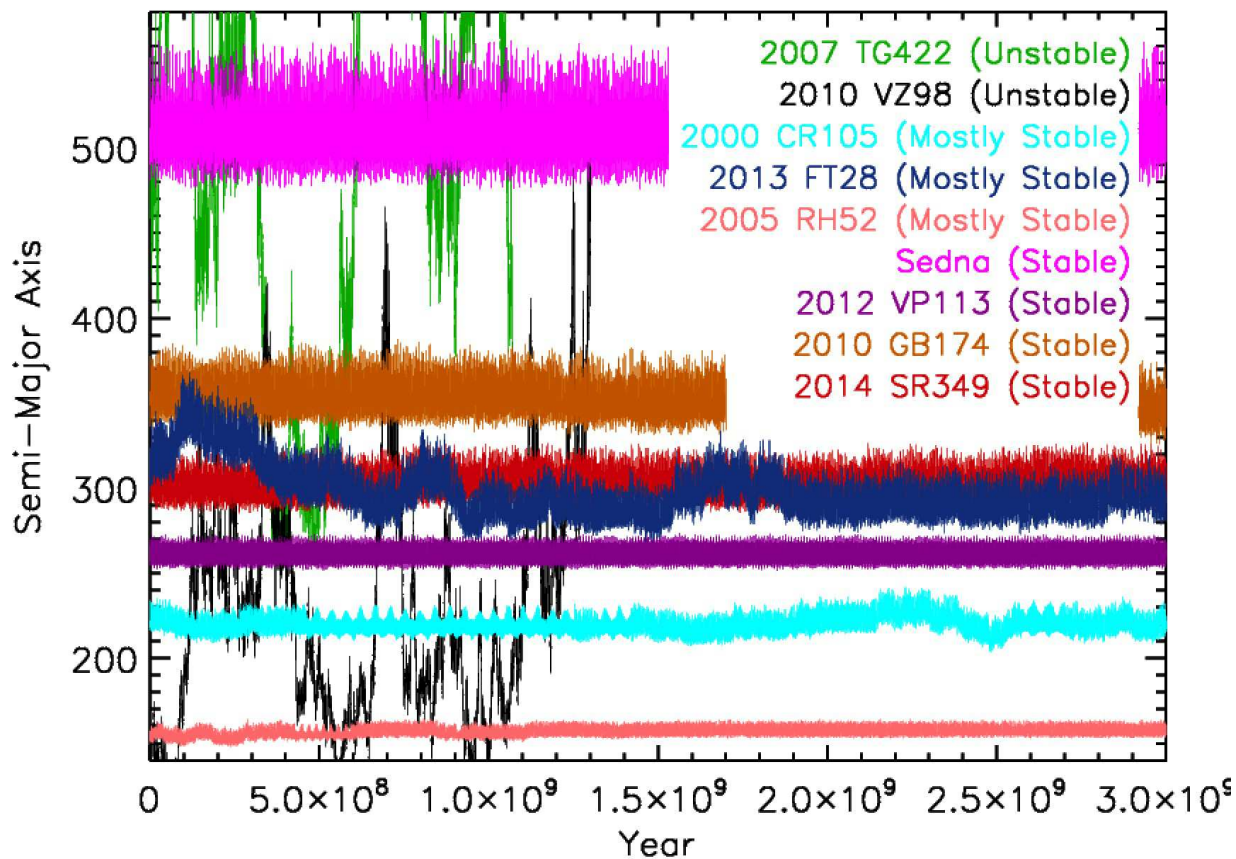}}
\caption{The numerically integrated nominal orbits of 2013 FT28,
  2014 SR349, 2012 VP113, 2010 VZ98, 2010 GB174, 2007 TG422, 2005 RH52,
  2000 CR105 and Sedna.  2014 SR349, like Sedna, 2012 VP113, and 2010
  GB174 is very stable over the age of the solar system.  2007 TG422
  and 2010 VZ98 appear to be very unstable and are lost within about 1
  billion years. 2013 FT28 is mostly stable, but shows some significant
  semi-major axis jumps suggesting resonant dynamical interactions
  with the giant planets.  2000 CR105 also is mostly stable but shows
  small semi-major axis spikes indicative of resonance interactions
  with the giant planets.  2005 RH52 also shows some semi-major axis
  spikes.  See Table 3 for a full list of ETNO orbit stability
  results.  It is clear the larger the semi-major axis of the stable
  extreme objects, the larger the range in semi-major axis the object
  has over the age of the solar system (a few AU for 2005 RH52 near
  150 AU and around 50 AU for Sedna near 500 AU).}
\label{fig:orbitsall2016} 
\end{figure}

It seems odd for BB2016 to have included 2010 VZ98 as this object has
one of the smallest semi-major axes and perihelia of the extreme
objects (Table 3).  We numerically integrated 2010 VZ98 and find it
doesn't pass their basic criteria as it shows significant semi-major
axis movement of over 100 AU within a few 100 Myrs
(Figure~\ref{fig:orbitsall2016}).  Thus as discussed above, the BB2016
statistical analysis should not have used 2012 VP113 as at that time
the discovery of 2012 VP113 was from a longitude of perihelion biased
survey.  It also appears they should not have used 2010 VZ98 as the
object has an unstable orbit (Table 3).  This would leave only four
extreme objects with stable orbits to test for longitude of perihelion
clustering, which would have given a non statistically significant
clustering result.  In this paper we have removed the longitudinal
biases from the discovery of 2012 VP113 and have found another stable
EDTNO in 2014 SR349, which brings the usable stable EDTNOs and IOC
objects back to six useful objects and a 0.7\% chance of being
clustered as observed in longitude of perihelion according to BB2016
statistics.

As we are using low number statistics, it is very important to
determine which ETNOs are useful and which are not when looking for
clustering.  To further understand the dynamics of the known ETNOs, we
numerically integrated all their nominal orbits for 4 billion
years.  In Table 3 we show the results of these numerical simulations.
As expected, the IOC objects 2012 VP113 and Sedna were found to be
very stable.  The most distant perihelion EDTNOs were also very
stable, 2010 GB174, 2014 SR349 and 2004 VN112.  2000 CR105 was stable,
but showed spiked motion in its semi-major axis indicative of possible
resonant interactions with the giant planets
(Figure~\ref{fig:orbitsall2016}).  This makes it questionable to use
2000 CR105 as a tracer for an unknown distant planet as significant
interactions with the interior giant planets might cloud the dynamics
of 2000 CR105.  We find the extreme scattered object 2005 RH52 is also
stable but again shows possible significant resonant interactions with
the giant planets from spiked semi-major axis motion.  We notice in
Figure~\ref{fig:orbitsall2016} that the larger the semi-major axis of
the stable extreme objects, the larger the range in semi-major axis
the object has over the age of the solar system ($\pm5$ AU for 2005
RH52 near 150 AU, $\pm8$ AU for 2012 VP113 near 260 AU, $\pm 15$ AU
for 2014 SR349 near 300 AU, $\pm20$ AU for 2010 GB174 near 370 AU, and
$\pm25$ AU for Sedna near 500 AU).

We found all extreme objects with perihelia less than 38.5 AU showed
semi-major axis variations greater than 100 AU over a 1 billion year
time-span, suggesting none may be good dynamical tracers of an unseen
distant planet.  Brown and Batygin (2016) further wrote about the
extreme objects in a second paper.  Here it is not obvious if they are
using 2010 VZ98, but their Figure 1 shows they are using 2007 TG422
and 2013 RF98, both of which are also found to be unstable in our
numerical integrations.  2013 RF98 has only a few month arc of
observations and thus has a very large orbital uncertainty, but worse
for 2013 RF98 is that its longitude of perihelion, though possibly
similar to the other stable ETNOs, is best not to use.  This is
because 2013 RF98 was found by the Dark Energy Survey Collaboration et
al. (2016) which has a strong longitude of perihelion angle bias since
the survey only observes on the side of the sky that Sedna and 2012
VP113 are in, where objects with these longitude of perihelion angles
would be brightest and easiest to find.  Brown and Batygin (2016)
appear most interested in ETNOs with $a>250$ AU and is the reason they
highlight 2007 TG422 and 2013 RF98.  We find 2013 RF98 and 2007 TG422
are unstable from Neptune perturbations over 10 Myr times scales, but
because of their very distant semi-major axes the hypothesized more
distant planet could have a much stronger effect on these orbits and
thus dominate over any Neptune or Uranus perturbations.  If this is
the case, than 2013 FT28 should also be used as it is much more stable
then these other objects.  If the clustered unstable objects are lost
on short time-scales, their must be a reservoir somewhere replacing
them.

\subsubsection{Statistics of ETNO Clustering}
It is important to determine which ETNOs are best to use in low number
statistical analyses.  18 of 19 extreme objects shown in Table 3,
stable or unstable, show the same clustering in argument of perihelion
between 290 and 40 degrees as first identified in TS2014.  Only 2013
FS28 does not show this clustering, but 2013 FS28 also has the second
lowest perihelion ($q\sim 34$ AU) of any ETNO and thus likely has
strong interactions with Neptune and thus we suggest only using
objects with $q>35$, though $q>40$ AU may be more appropriate to
further remove Neptune interactions.  We further have shown that there
is little to no bias in the clustering of the argument of perihelion
from discoveries and have shown that strangely, the objects with
$a>200$ AU and $q<35$ AU show a 180 degree opposite clustering for the
argument of perihelion angles (100 to 200 degrees).  Finally, we also
showed that there is a correlation between the argument of perihelion
and longitude of perihelion for ETNOs.  The unstable objects seem to
demonstrate the same clustering properties as the stable objects. This
suggests that the unstable objects also provide clues to the dynamical
processes in the 50 AU to 1500 AU outer solar system regime and that
the unseen planet's interactions might dominate over Neptune
interactions.

For the longitude of perihelion clustering 12 of the 19 extreme
objects in Table 3 show clustering in the main longitude of perihelion
clustering between 0 and 130 degrees.  This is less distinct than the
clustering in argument of perihelion and thus one might want to limit
the ETNOs used to see if the clustering is stronger or weaker.  BB2016
talk about using only stable objects or objects with higher semi-major
axes of 250 or 350 AU instead of 150 AU.  Using only these objects,
the statistics get better, but the number of objects is low (6 of 9
for mostly stable ETNOs, 5 of 6 for completely stable ETNOs, 7 of 8
for $a>250$ ETNOs, 5 of 5 for completely stable, $a>250$ AU and $q>45$
AU ETNOs).  In the most stringent case there are only 5 useful objects
and thus the statistics are almost but not quite significant at the 3
sigma level.  Finding more of these ETNOs with very high perihelia
$q>45$ AU and semi-major axes $a>250$ AU is required to fully confirm
that the longitude of perihelion clustering is real.  The longitude of
perihelion clustering also gets a little uncertain since Brown and
Batygin (2016) showed objects could be in a secondary anti- or 180
degree opposite clustering in longitude of perihelion, which we
believe 2013 FT28 is part of.  With this, the range of acceptable
longitude of perihelia for ETNOs becomes much wider and thus harder to
look for significant outliers, but including both the primary and 180
degree secondary longitude of perihelion angles, we find for $a>250$
AU that 8 of 8 objects are clustered in longitude of perihelion.

Clustering in the Longitude of Ascending node for high perihelion
objects is also apparent for the two different longitude of perihelion
clusters (Figure~\ref{fig:kboaLong2016more}).  Our numerical
simulations show that the argument of perihelion cycles through all
angles about twice as fast as the longitude of perihelion angle for
each extreme object when only including the known major planets.  As
noted earlier, the argument of perihelion appears to be significantly
correlated with the longitude of perihelion for all ETNOs.

\subsection{Likelihood of Orbital Clustering Through An Observational Bias Simulator}

In section 3 above we described the basics of our observational
simulator.  In our real observational survey data, we found no IOCs
nor EDTNOs with $40\deg < \omega < 290\deg$ nor any with $230\deg <
\Omega < 360\deg$ and 1 IOC and 2 EDTNOs outside of these ranges.  We
find the odds of detecting 1 IOC and 2 EDTNOs in the $290\deg < \omega
< 40\deg$ regime is 10\% from our survey in the radially symmetric
case. The odds of detecting 1 IOC and 2 EDTNOs with $0\deg < \Omega <
230\deg$ in the radially symmetric case is 18\%. Assuming $\Omega$ and
$\omega$ are statistically uncorrelated results in a 1.8\% probability
that this could happen by chance from our survey detections of the IOC
and 2 EDTNOs alone if the populations are radially symmetric.

Both Sedna and 2012 VP113 were found in similar sky locations as
reported in TS2014.  Sedna was discovered at a right ascension of
03:13 hours and a declination of +05:47 degrees and 2012 VP113
discovered at a right ascension of 03:23 hours and a declination of
+01:12 degrees, less than 6 degrees apart.  We quantified how many
IOCs would be detected in our simulation within 6 degrees of Sedna's
discovery location given our observed fields and survey
methodology. The probability of this occurring by chance in a
symmetric population given our survey field distribution is less than
1 in 1000. Thus, this provides further evidence that the IOCs are in
an longitudinally asymmetric population as first suggested in TS2014.

We test for observational evidence for longitude of perihelion by
simulating a population that is uniform in terms of the angular
elements $\omega$ and $\Omega$ (and thus longitude of perihelion
$\bar{\omega}=\omega + \Omega$). Then we estimate the probability that
the true distribution of objects violates the discovery statistics
expected under these assumptions. Given the uniform $\omega$ and
$\Omega$ case, one would expect 35 of 130 detected simulated objects
to fall in the $0 < \bar{\omega} < 130\deg$ range given our true
detection statistics. In our survey, 2 extreme objects fell into this
range, 2012 VP113 and 2014 SR349 and one outside 2013 FT28.  If we
reject 2013 FT28 from this sample because it may have significant
Neptune interactions or because it may lie in the secondary
anti-longitude of perihelion clustering, the probability of this
longitude of perihelion range happening by chance in our survey alone
is 7\%, which is not statistically significant.

\subsubsection{Biases Of Other Discovered Extreme TNOs}
Other surveys which have detected extreme TNOs may have similar biases
to our survey if they have covered a large amount of sky generally
distributed throughout the year in terms of search date.  Under this
assumption, there are 5 known stable IOCs and EDTNOs with $a>250$ AU
that fall into the $0 < \bar{\omega} < 130\deg$ range (2012 VP113,
Sedna, 2010 GB174, 2014 SR349 and 2004 VN112: Table 3).  The probability
of all 5 known objects having a clustered $\bar{\omega}$ between 0 and
130 degrees by chance is then $(35/130)^5 = 1.4 \times 10^{-3}$ which
is around $3\sigma$ assuming Gaussian statistics.

But other surveys that have found extreme TNOs do not have the minimal
biases as our survey, which makes the assumption of similar biases
untrue.  As detailed earlier, 2012 VP113 and 2014 SR349, found in this
work, have minimal to no bias in longitude of perihelion.  Sedna,
found in a mostly all sky survey, also likely has no significant
longitude of perihelion bias (Brown et al. 2004).  But 2004 VN112 was
found in the ESSENCE supernova survey that was mostly obtained between
October and December, which is similar to the location on the sky of
Sedna and 2012 VP113 (Becker et al. 2008).  So 2004 VN112 has a strong
longitude of perihelion bias, though should have no significant
argument of perihelion bias as it was a survey well off the ecliptic.
2010 GB174 was found in a small limited survey well off the ecliptic
that was focused near the Virgo cluster around 12 to 13 hours in RA
(Chen et al. 2013).  This is on the opposite side of the Milky Way
than Sedna and 2012 VP113, but the limited survey area makes it
questionable how to handle the bias in longitude of perihelion for
2010 GB174.

Some of the other objects shown in Table 3 that we find are more
unstable do have minimal longitude of perihelion discovery biases.
2000 CR105, which was found in the Deep Ecliptic Survey (DES) that
observed at all times of the year (Elliot et al. 2005), 2005 RH52 from
the CFHT Ecliptic Plane Survey (Petit et al. 2011) and 2013 GP136
found in the Outer Solar System Origins Survey (Bannister et
al. 2016).  As mentioned earlier, 2013 RF98 has a strong longitude of
perihelion bias as the Dark Energy Survey that found it only observes
the sky near Sedna and 2012 VP113 (Dark Energy Collaboration et
al. 2016).  Most of the other objects shown in Table 1 do not have
known discovery biases.

Thus with the new data reported here, it appears that the longitude of
perihelion clustering for the ETNOs is real and there are two distinct
anti-clusters, as long as assumptions about other survey biases not
being important are true, stable and unstable objects are equally
reliable tracers, and one places 2013 FT28 into the secondary
anti-longitude of perihelion clustering region.  Further unbiased
extreme TNO discoveries need to be made to confirm the longitude of
perihelion clustering.

\subsection{Extreme TNOs Orbital Period Ratios}

Malhotra et al. (2016) found that four of the EDTNOs (Sedna, 2010
GB174, 2004 VN112 and 2012 VP113) would have N/1 or N/2 period ratios
with a hypothetical massive planet with a semi-major axis around 665
AU.  These period ratios could be an indication of the EDTNOs being in
some sort of mean motion resonance with a distant massive planet.  The
two new ETNOs reported here, 2014 SR349 and 2013 FT28, have fairly
similar semi-major axes, eccentricities and inclinations, though they
have opposite longitude of perihelia as discussed above.  2014 SR349
and 2013 FT28 would have about 4880 and 5460 year orbital periods and
are not similar to any of the other known EDTNOs.  The new EDTNOs
would have an orbit period ratio of about $3.5\pm0.3$ and
$3.13\pm0.1$ respectively, if the hypothetical unknown distant
massive planet had a period of 17116 years.  Thus 2014 SR349 could be
near the 7/2 resonance, though the uncertainty in the orbit is still
too large to make any definitive statements.  2013 FT28 has a better
constrained orbit and isn't near any of the major resonances, though
the 3/1 isn't too far off.

\section{Population Statistics}
\subsection{Extreme Detached TNOs Population Statistics}

We considered two possible EDTNO populations: one with a uniform
distribution of the angular orbital elements argument of perihelion
and longitude of ascending node, and a second with a constrained set
of angular orbital elements noted by TS2014 and BB2016.  For the
constrained angular case, we used an argument of perihelion range of
$-80\deg < \omega < 45\deg$ and a longitude of perihelion range of
$0\deg < \bar{\omega} < 130\deg$. The longitude of ascending node was
then computed from $\Omega = \bar{\omega} - \omega$.

\begin{figure}
\epsscale{0.4}
\centerline{\includegraphics[angle=0,totalheight=0.55\textheight]{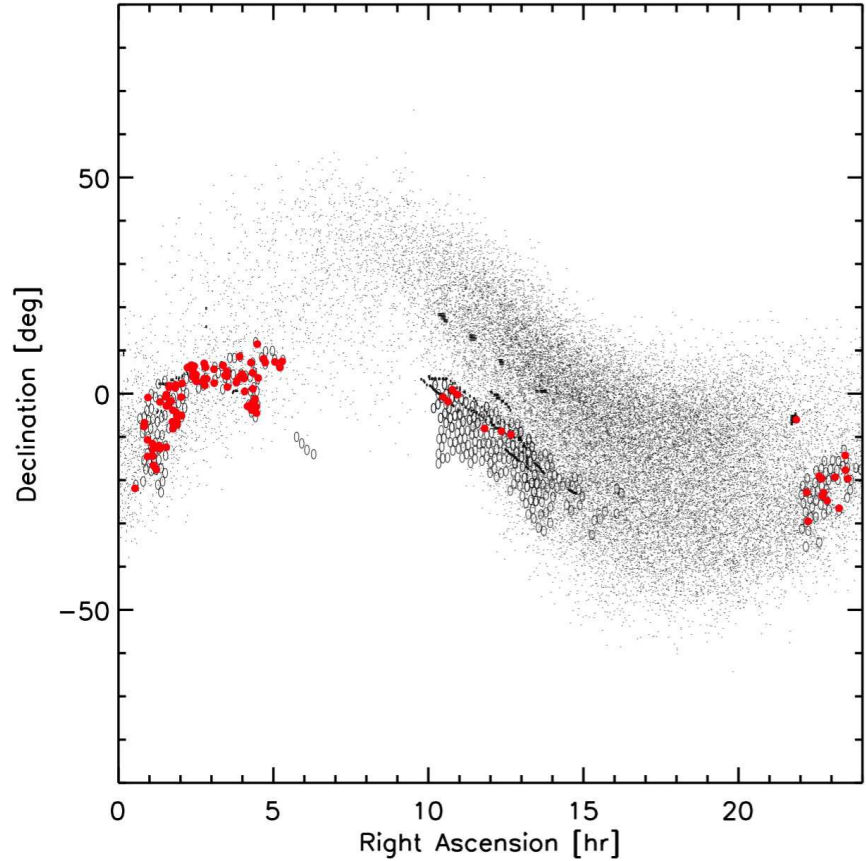}}
\caption{Our survey fields (open circles) shown with simulated
  discovered objects (red) from our survey simulator.  Only a fraction
  of the objects in our simulation are shown as small black dots.
  This simulation was based on the extreme trans-Neptunian objects as
  having an asymmetric orbital distribution where their longitudes of
  perihelion are all near Sedna and 2012 VP113.  Thus most of the
  objects would be located near aphelion around 14 to 19 hours in
  right ascension while most discoveries would be made when the
  objects are brightest near perihelion between 2 and 7 hours.  This
  figure shows the bias a survey would have in discovering extreme
  objects if only obtained near one particular longitude.}
\label{fig:q55RaDecSim} 
\end{figure}

\begin{figure}
\epsscale{0.4}
\centerline{\includegraphics[angle=0,totalheight=0.6\textheight]{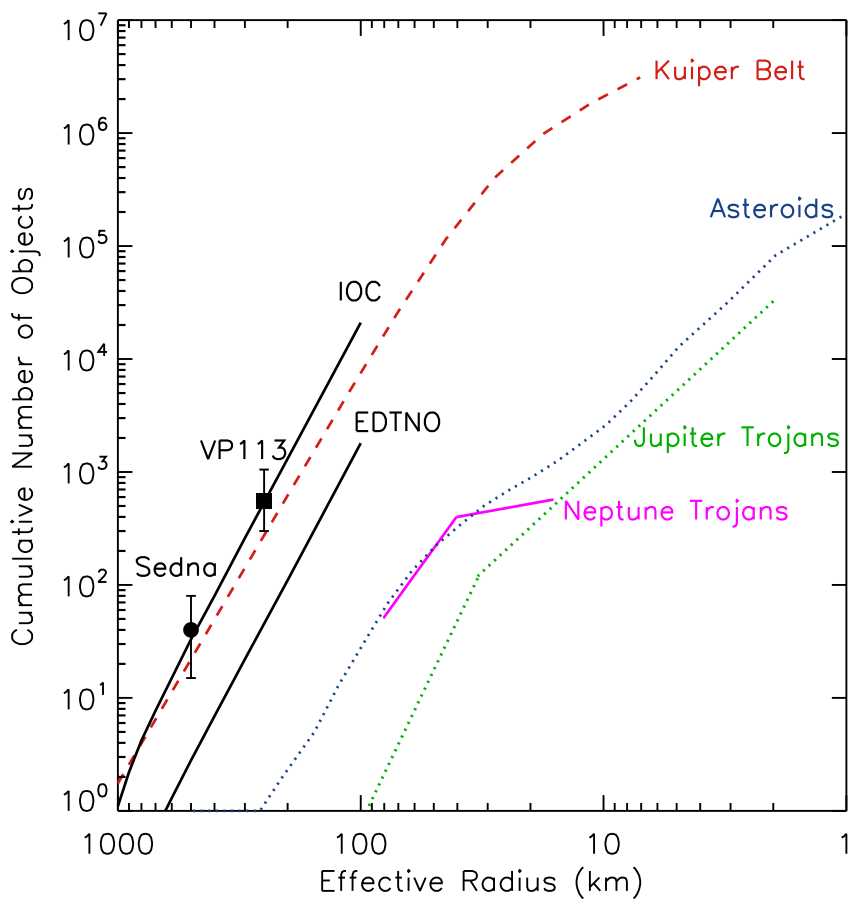}}
\caption{The cumulative number of objects in the known stable
  reservoirs in our solar system. The IOC population ($q>50$ AU) is
  likely the largest observed reservoir of small objects in our solar
  system and should have a few objects larger than Pluto within the
  population. The errors on the population are shown by the 2012 VP113
  and Sedna locations.  The extreme detached trans-Neptunian objects
  ($40<q<50$ AU) are likely smaller than the IOC population.  Both the
  IOC and EDTNO population estimates are based on our survey and is
  explained in the text.  (Population estimates are from This Work,
  Jewitt et al. 2000; Jedicke et al. 2002; Bottke et al. 2005; Fraser
  \& Kavelaars 2008; Fuentes \& Holman 2008; Sheppard \& Trujillo
  2010).}
\label{fig:allsize2016} 
\end{figure}

To estimate the population of EDTNOs we used a radially symmetric and
asymmetric population described in Table 5 based on our 2 detections
of EDTNOs (2014 SR349 and 2013 FT28) and the results of our observational
bias simulator (Figure~\ref{fig:q55RaDecSim}). For the asymmetric
case, we find that the total number of EDTNOs in the radius range
$100\mbox{km} < r < 4000 \mbox{km}$ to be 1800 and the total mass to
be $3.8 \times 10^{22}$ kg ($\sim 1/150$ of an Earth mass) assuming a
differential power law of $q' = -5$
(Figure~\ref{fig:allsize2016}). Due to the number of assumptions in
our simulation and the small number of detections, these quantities
are only order of magnitude estimates. As more EDTNOs are discovered
in our ongoing survey, we will be able to pin down the total mass of
the EDTNOs with much more accuracy. Population estimates for other
assumptions, such as $q' = -4$ and symmetric populations, can be found
in Table 5.

\subsection{Updated Inner Oort Cloud Object Population Statistics}

In TS2014 we gave IOC population statistics assuming a symmetric
population, but stated in TS2014 that 2012 VP113 was found at a
similar sky position as Sedna and thus the population might not be
longitudinally symmetric on the sky (i.e. there could be longitude of
perihelion clustering).  The IOC population would be smaller than that
reported in TS2014 if the IOC objects are not uniformly distributed
across the sky.  This survey has now covered almost 20 times more sky
area than analyzed in TS2014 and thus the population statistics of the
IOC objects is likely significantly less than reported in that paper
as no new IOC object has been found in a much more uniform sky survey.
Our survey as reported in TS2014 was longitudinally biased and thus it
was not useful to determine if the IOCs where longitudinally
asymmetric.  Our much larger survey reported here is not
longitudinally biased and thus we can report revised population
statistics for the IOCs assuming an asymmetric population in
longitude.

Our current survey found one inner Oort cloud object ($q>50$ AU and
$a>150$ AU), 2012 VP113.  For our model populations we restricted our
minimum perihelion to 75 AU since the two known IOCs, Sedna and 2012
VP113, both have perihelia above this range, which might suggest an
inner edge around 75 AU (Trujillo and Sheppard 2014).  We assumed that
the IOCs also have constrained angular orbital elements, similar to
the EDTNOs (Table 5). From the detection of 2012 VP113 and the survey
fields we have covered, we find the number of bodies in the radius
range $100\mbox{km} < r < 4000 \mbox{km}$ to be $2 \times 10^4$ and a
total mass of $4 \times 10^{23}$ kg or $\sim 1/10$ of an Earth mass
(Figure~\ref{fig:allsize2016}).  We expect there should be a few
objects larger than Pluto in the IOC.  These numbers are very
dependent on the eccentricity distribution of the IOCs. For instance,
we are very insensitive to the most eccentric IOCs since they spend
most of their time at aphelion and too faint to be detected. Thus, the
eccentricity distribution and semi-major axis distribution have a
large impact on the total mass and number of bodies estimated. We have
chosen fairly conservative numbers for these, so if anything, the
estimated population of IOCs is likely to increase with further
observational constraints rather than decrease.

\subsection{The Inner Oort Cloud Inner Edge}

One important question is whether the EDTNOs and the IOCs could be
drawn from the same population. Currently, the two known IOCs Sedna
and 2012 VP113 have perihelia greater than 75 AU. In contrast, the
EDTNOs all have perihelia less than 50 AU. The main observational
question is then whether a single set of population-wide parameters
could encompass both populations, or whether there is a gap or dearth
of objects in the 50 AU to 75 AU regime. This was simulated by TS2014
who determined that the paucity of objects in the 50 AU to 75 AU
regime was most likely not observational bias using both the original
survey that found Sedna and their survey. We repeat this simulation
here for only the survey described in this paper. This is a more
thoroughly characterized survey and represents and additional factor
of $\sim 20$ more sky coverage than our survey in TS2014.

If the EDTNOs and IOCs were drawn from the same population, one would
expect large semi-major axis objects to be found with perihelion in
the 50 AU to 75 AU regime. In our survey simulator, we find that 71\%
of detections should occur in the 50 AU to 75 AU range. Since we found
2 EDTNOs and 1 IOC object, none of which had perihelia in the 50 to 75
AU range, the probability of this occurring by chance is about 2\%. It
is not clear whether including the extreme scattered objects 2013 UH15,
2013 FS28 and 2014 SS349 would be appropriate, but if they were included
the probability would drop to 0.1\% and thus be statistically
significant above the $\sim 3\sigma$ level using our survey data
alone.

Other surveys have found the other 19 known ETNOs, none of which have
perihelion in the 50 AU to 75 AU range. If other surveys have similar
biases to our own, the probability of this occurring by chance is
roughly $0.29^{19} = 6 \times 10^{-11}$ or about 7 $\sigma$ assuming
Gaussian statistics.

\subsection{Cumulative Luminosity Function}

\begin{figure}
\epsscale{0.4}
\centerline{\includegraphics[angle=0,totalheight=0.6\textheight]{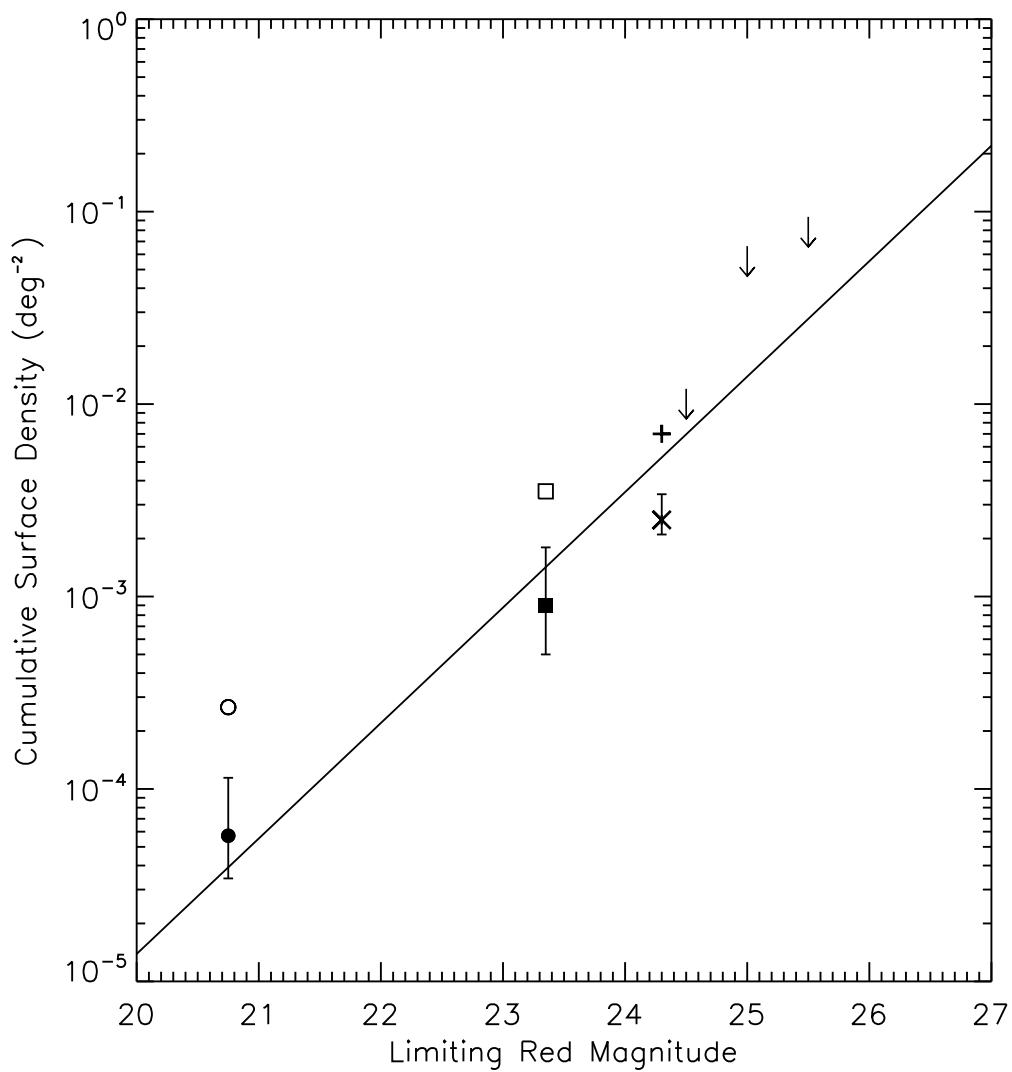}}
\vspace{0.5in}
\caption{The Cumulative Luminosity Function for Inner Oort Cloud
  Objects in the r-band.  Sedna is shown by the filled circle while
  2012 VP113 is shown by the filled square.  For null
  detections, the three sigma upper limits are shown with arrows.  The
  solid line is the Kuiper Belt luminosity function shifted downwards
  to match the two known IOC detections.  The X shows the two extreme
  detached trans-Neptunian objects (2014 SR349 and 2013 FT28) along with
  the 2012 VP113 detection from our survey.  This result is based on
  uniform sky coverage of IOC objects.  If they are clustered in
  longitude of perihelion between about 2 and 7 hours in right
  ascension, their cumulative surface densities would hold only in
  that region of sky and be increased by over a factor of four as
  shown by the open symbols.  The plus shows just 2014 SR349 and 2012
  VP113 assuming clustered longitude of perihelion as 2013 FT28 would
  not be found in this region.}
\label{fig:sednacum2016} 
\end{figure}

The Cumulative Luminosity Function (CLF) describes the sky-plane
number density of objects brighter than a given magnitude.  The CLF
can be described by
\begin{equation}
\mbox{log}[\Sigma (m_{r})]=\alpha (m_{r}-m_{o})  \label{eq:slope}
\end{equation}
\noindent where $\Sigma (m_{r})$ is the number of objects brighter
than $m_{r}$, $m_{o}$ is the magnitude zero point, and $\alpha$
describes the slope of the luminosity function.  There is no known CLF
for the IOC objects or EDTNOs as few are known.  We here attempt a
crude estimate based on our detection of one IOC object (2012 VP113)
and two EDTNOs (2014 SR349 and 2013 FT28).  The CLF for these objects is
shown in Figure~\ref{fig:sednacum2016}.  The slope for the CLF for the
IOC objects appears consistent with that found for the moderate sized
Kuiper Belt objects ($\alpha \sim 0.6$: Fraser \& Kavelaars 2008;
Fuentes \& Holman 2008), but just shifted downwards ($m_{o} \sim 28$)
to adjust for the lower number of objects per brightness bin on the
sky.

\section{Additional Interesting Objects Discovered}

\subsection{Outer Oort Cloud Discovery}
The first object that enters into the outer Oort cloud yet has
perihelion greater than Neptune was discovered as part of this
survey. 2014 FE72 has a semi-major axis around 2155 AU and an aphelion
distance of some 4000 AU.  Thus outside forces such as the galactic
tide and stellar perturbations have likely influenced the orbit of
2014 FE72 over time (Duncan et al. 2008; Soares and Gomes 2013; Kaib
and Quinn 2009).  The moderate perihelion of 2014 FE72 around 36 AU
suggests that its highly elliptical orbit is from past Neptune
encounters, but its longitude of perihelion $\bar{\omega}=111$ degrees
is similar to the other known high semi-major axes objects
($0<\bar{\omega}<130$ degrees).  This could suggest that the possible
massive unknown outer planet is shepherding 2014 FE72 similarly as the
other EDTNOs and IOCs.  Since 2014 FE72's inclination is only around
20 degrees, it is similar to the other extreme objects.

\subsection{The Curious Case of 2014 SS349}
The object 2014 SS349 is a borderline extreme object as it has a
semi-major axis just below the ETNOs cutoff of 150 AU.  The strength
of a Neptune resonance decreases as the semi-major axis of an object
increases and thus the large semi-major axes of the ETNOs and IOC
objects suggest they are not strongly perturbed by resonances from
Neptune and this may be the case of 2014 SS349 as well.  2014 SS349 has a
fairly large inclination of 48 degrees, suggesting its orbit is
modified by the Kozai resonance.  Though its perihelion is now about
45.5 AU, it could travel much closer to Neptune's orbit through the
Kozai mechanism alone or in combination with some Neptune mean motion
resonance interaction in the past (see Gomes et al. 2008 or Kaib \&
Sheppard 2016).  Thus 2014 SS349 is probably more like the high
perihelion objects with moderate eccentricities and semi-major axes
discussed in Sheppard et al. (2016) that likely have a different
origin and history than the ETNOs.

\subsection{New Extreme Scattered Disk objects}
The objects 2013 UH15 and 2013 FS28 have large semi-major axes, but their
perihelia are too close to Neptune ($q<35$ AU) to be considered EDTNOs
like 2014 SR349 and 2013 FT28.  Both objects were lost within the age of
the solar system during numerical integrations through Neptune
interactions, so the most likely explanation for these extreme
scattered objects is that they are related to scattered disk objects
like 1996 TL66 (Luu et al. 1997; Duncan and Levison 1997; Gomes et
al. 2008).  The new discovery 2013 FS28 is the only ETNO of 19 known
that does not follow the argument of perihelion clustering for objects
with $a>150$ AU and perihelia greater than Neptune.  2013 FS28 has the
second lowest perihelion of any ETNO so likely has strong Neptune
interactions.

\subsection{Possible High Order Neptune Mean Motion Resonances}
2014 SW349, 2014 FJ72 and 2014 FL72 appear to have similar orbits (Table
4).  Their moderately high perihelia of 38 to 39 AU makes them
significantly far from Neptune to have strong Neptune interactions.
They all could be in some high order mean motion resonance with
Neptune, or could just be typical scattered disk objects.  Though they
have relatively high perihelia, they are not extreme because of their
moderate semi-major axes.  All three objects appear stable for the age
of the solar system in our numerical simulations.

Objects 2014 SZ349 and 2014 SD350 are near the 13:3 Neptune mean motion
resonance. Both have perihelia fairly close to Neptune, which allows
Neptune to strongly influence their orbits.  Neither 2014 SD350 or
2014 SZ349 were stable for the age of the solar system in our numerical
simulations.

\subsection{Other Discoveries}
This survey has discovered some of the most distant objects ever
observed in the solar system.  The TNO 2013 FY27 was found to be one
of the most distant and one of the brightest outer solar system
objects discovered in this survey.  The 22 magnitude r-band magnitude
and 80.5 AU distance of 2013 FY27 give the object an absolute
magnitude of about 2.9.  This makes 2013 FY27 one of the top ten
intrinsically brightest TNOs and thus it could be a top ten largest
TNO as well. Though the diameter and albedo are unknown, assuming an
albedo of 0.1 yields a diameter around 1000 km.

A new possible bright Haumea family member was discovered as well,
2014 FT71 ($a=43.8$ AU, $e=0.14$, and $i=27.9$ deg), based on orbital
similarity to other known members of the Haumea family (Ragozzine and
Brown 2007). If true, infrared spectroscopy of this object should show
very strong water ice absorption (Brown et al. 2007; Snodgrass et
al. 2010; Trujillo et al. 2011).

One well know main belt asteroid, (62412) 2000 SY178, was found to
show a tail during our survey observations (Sheppard and Trujillo
2015).  This object is now classified as an active asteroid (Jewitt
2012; Hsieh et al. 2015) and was found to be part of the Hygiea outer
main belt asteroid family with a fast rotation period.  Further
details about active asteroid 62412 are available in Sheppard and
Trujillo (2015).  Through the survey, comet C/2014 F3
(Sheppard-Trujillo) was discovered to be one of the most distant
active comets ever observed at 13 AU.  Its perihelion is near Jupiter,
suggesting strong interactions with Jupiter.  Finally, 2007 TY430 was
discovered and found to be an ultra-red, equal-sized, wide binary
Kuiper Belt object in the Neptune 3:2 mean motion resonance.  This
makes 2007 TY430 the only known ultra-red, equal-sized, wide binary
known outside of the main classical Kuiper Belt (see Sheppard et
al. 2012).

For completeness, we here list all the objects discovered beyond 50 AU
during our survey to date.  The most updated orbits for the objects
can be found at the Minor Planet Center.  For 2012 VP113 see Trujillo
and Sheppard (2014).  For 2014 FZ71, 2013 FQ28, 2015 GP50, 2014 FC69
and 2012 FH84 see Sheppard et al. (2016).  Other discoveries beyond 50
AU with decent orbital elements are: 2013 FY27, 2013 FT28, 2013 FS28,
2013 UH15, 2014 FE72, 2014 SR349, 2014 SS349, 2014 FJ72, 2014 FL70,
2011 GM89, 2012 FN84, 2012 OL6, 2013 UJ15, 2013 UK15, 2014 FF72, 2014
FL72, 2014 FM72, 2014 FH72, 2014 FG72, 2014 FK72, 2014 SJ350, 2014
SV349, 2014 SG350, 2014 SK350, 2014 SD350, 2014 SW349, 2014 SY349,
2014 SX349, 2014 SB350, 2014 SL350, 2014 SE350, 2014 SC350, 2014
SM350, 2014 SF350, 2014 ST349, 2014 SU349, 2014 SZ349, 2015 GR50, 2015
GS50, and 2015 GQ50. A few objects were not followed after about 1
month of observation arc as their orbits didn't look extreme or
interesting or were found to be at less than 50 AU distance after
recovery: 2012 FM84, 2013 VJ24, 2014 FN72, 2014 FP72, 2014 FO72, 2014
SA350, and 2014 SH350. Finally, there were two objects that were in
the noise at discovery and appear to be beyond 50 AU, but were too
faint to easily recover: D359c02 (found on field decam359 on chip 02)
around 24.5 mags near 65 AU at 13:24:11 hours and -13:26:48 degrees on
March 17.283, 2013 and D364c08 24.5 mags around 70 AU at 13:24:51
hours and -24:20:08 degrees on March 17.301, 2013.

\section{Finding a Super-Earth to Neptune Mass Planet in the Outer Solar System}

\begin{figure}
%\epsscale{0.4}
\centerline{\includegraphics[angle=0,totalheight=0.6\textheight]{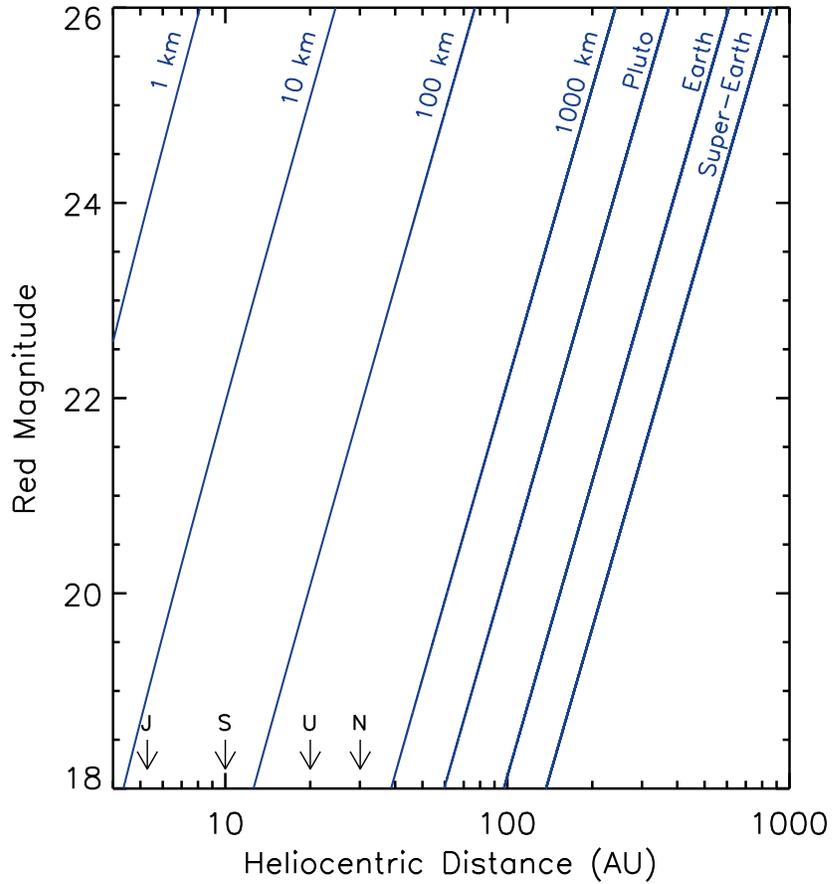}}
\caption{The r-band magnitude of different sized objects and
  heliocentric distances assuming a moderate albedo of 0.20.  A
  super-Earth of about 10 Earth masses would have about twice the
  radius of Earth.  The biggest telescopes with wide-field imagers can
  efficiently survey to just below 26th magnitude.  Thus a Super-Earth
  beyond several hundred AU would likely be fainter than any current
  wide-field survey capability.}
\label{fig:distance26log2016} 
\end{figure}

We also simulated our survey's sensitivity to finding the distant
massive planet predicted by TS2014 with the recently hypothesized
rudimentary orbit described by BB2016.  Our survey has attempted to
obtain uniform sky coverage in both ecliptic latitude and longitude to
prevent biases in the orbital parameters of discovered objects.  To
this end, we find that our survey would not detect such a distant
eccentric planet some 99\% of the time. This is mostly because the
possible hypothesized planet's large eccentricity implies that it
spends most of its time near aphelion, which would make it very faint
and a significant fraction of its time in the northern sky if it is
anti-aligned with the ETNOs. Most of the sky area covered in this work
was in the southern sky due to our use of DECam at CTIO, which has a
latitude near 30 degrees South. Our continuing survey is now covering
large portions of the Northern sky using HyperSuprimeCam on the Subaru
telescope and will be detailed in future work.
Figure~\ref{fig:distance26log2016} shows that even a very massive 10
Earth mass planet beyond a few hundred AU would likely go undetected
in most surveys to date as its r-band magnitude would be fainter than
24th magnitude with moderate to dark albedo.

In this work we have used IOC to refer to bodies with extremely high
perihelia ($q>75$ AU) for consistency with past works.  We note that
if the distant massive planet hypothesis is correct, it implies that the
IOC population is not in fact primordial, nor is it directly related
to the Oort cloud. What we have been calling the IOC would actually be
a highly evolved dynamical population that is strongly influenced by
the distant massive planet.  Thus, a different nomenclature should be
used when discussing bodies like Sedna and 2012 VP113. Calling IOC
objects as just simply extreme detached objects seems most logical as
it would appear both these populations would be directly related in
the Planet X/Planet 9 regime.  Since the unseen planet has not been
discovered and no such nomenclature has been agreed upon, it is an
open discussion.  Calling the objects most perturbed by the putative
distant planet Sednoids could also be logical after the brightest and
first obvious member of this population.

\section{Summary}

We have covered about 1080 square degrees of sky to over 24th
magnitude searching for objects beyond the Kuiper Belt edge.  Unlike
most faint outer solar system surveys, our survey has fairly uniform
sky coverage in ecliptic longitude and southern ecliptic latitudes
between about 5 and 20 degrees.  Thus our survey should have little to
no bias in the longitude of perihelion of discovered objects.  The
average ecliptic latitude was some 13 degrees from the ecliptic plane,
which would limit the argument of perihelion bias during discovery
that previous ecliptic plane surveys could have.  Several new extreme
objects with perihelia greater than 40 AU and semi-major axes greater
than 150 AU were discovered.

1) Newly discovered extreme detached trans-Neptunian object 2014 SR349
has a semi-major axis around 288 AU and a distant perihelion of almost
48 AU.  2014 SR349 was found to have a stable orbit over the age of
the solar system and a argument of perihelion and longitude of
perihelion similar to the other known extreme trans-Neptunian objects
and inner Oort cloud objects.  Thus 2014 SR349 continues to show the
orbital clustering of extreme trans-Neptunian objects first identified
and explained as likely caused by a distant massive planet between
about 200 to 1500 AU in the outer solar system in Trujillo and
Sheppard (2014).

2) 2013 FT28 is also a newly found extreme detached trans-Neptunian
object with a similar semi-major axis as 2014 SR349 and slightly lower
perihelion of 43.6 AU.  The argument of perihelion for 2013 FT28 is
consistent with the clustering of argument of perihelia for extreme
objects, but its longitude of perihelion is not as it is about 180
degrees away from the other extreme detached objects.  2013 FT28
appears to be the first high semi-major axis object within the
secondary longitude of perihelion clustering that could be 180 degrees
from the main clustering discussed in Brown and Batygin (2016) for an
eccentric, inclined distant massive planet.  2013 FT28 is the highest
perihelion object with a semi-major axis above 250 AU to show some
orbit instability and thus demonstrates significant Neptune
interactions are possible even for objects with perihelia around 43.6
AU.

3) The argument of perihelion clustering for extreme objects is
strongly statistically significant at about the $6 \sigma$ level with
all 15 objects with $q>35$ and $a>150$ AU having arguments of
perihelion between 290 and 40 degrees.  In addition to the extreme
objects with perihelia greater than 35 AU having argument of perihelia
clustered in the range between about 290 and 40 degrees, we find the
opposite is true for objects with perihelia less than 35 AU and $a>
200$ AU.  These high semi-major axis trans-Neptunian objects with
perihelia less than 35 AU have arguments of perihelia that cluster in
the range 100 to 200 degrees.  This reverse clustering is not yet
explained.

4) The longitude of perihelion clustering is not as robust as the
argument of perihelion clustering.  Past statistical analyses on the
possible longitude of perihelion clustering of extreme trans-Neptunian
objects used objects with biased longitude of perihelion discoveries.
With the new results from this work, we show the longitude of
perihelion clustering is likely real.  With the discovery of 2013
FT28, we find there are two longitude of perihelion clusters for
extreme objects about 180 degrees apart.

5) The longitude of perihelion is significantly correlated with the
argument of perihelion at the 99.99\% level for ETNOs as those with
$\bar{\omega}=0$ to 130 have $\omega= 280$ to 360 and come to
perihelion below the ecliptic while those with $\bar{\omega}=180$ to
340 have $\omega= 0$ to 40 degs and come to perihelion above the
ecliptic.  We find these two groups also have correlated longitude of
ascending node clustering and orbit pole angle clustering.  Thus we
find two distinct ETNO clustering groups that are anti-aligned with
each other.  This correlation is further evidence of an unknown
massive planet in the outer solar system.  This is because extreme
eccentric objects with perihelia on opposite sides of the sky (180
degree longitude of perihelion differences) would approach the planet
at opposite points in their orbits.  Thus for the extreme objects to
stay away from the planet, they would prefer opposite ecliptic
latitudes of the planet in order to avoid the planet (i.e. opposite
argument of perihelia or orbit pole angles).

6) 2014 SS349 is an object with a high perihelion ($q=45.5$ AU), fairly
high semi-major axis ($143$ AU) and high inclination ($i=48$ degrees).
It is a borderline extreme detached object like 2014 SR349 and 2013 FT28,
but its much higher inclination and lower semi-major axis suggests it
is likely to be much more influenced by the Kozai resonance.  The
dynamics of this object need to be further explored, but it is not
considered an extreme object.

7) 2014 FE72 has a semi-major axes around 2155 AU and a aphelion around
4000 AU and likely experiences significant perturbations from outside
forces such as the galactic tide.  This makes 2014 FE72 the first outer
Oort cloud object discovered that has a perihelion beyond Neptune.
This object likely interacted with Neptune in the past to obtain such
a distant eccentric orbit.  Surprisingly, the longitude of perihelion
of 2014 FE72 is similar to the other ETNOs with very large semi-major
axes.

8) The cumulative luminosity function for the inner Oort cloud objects
and extreme objects suggests a similar slope as the Kuiper Belt object
population.  Based on the area and objects discovered in this survey,
we estimate about 21000 inner Oort cloud objects ($q>75$ AU) and 1800
extreme detached trans-Neptunian objects ($40<q<50$ AU) exist larger
than 100 km in size assuming an asymmetric population with most coming
to perihelion between about 2 and 7 hours in Right Ascension.

9) The inclinations of the EDTNOs and IOCs are moderate with all but
Sedna being between 17 and 26 degrees.  Their average inclination is
about 20 degrees, which is very similar to the scattered disk
population.

10) We note that if there is a massive planet shepherding the inner Oort
Cloud objects and extreme detached objects in their orbits, these two
populations are likely to be from the same source.  Further, the
orbits of the inner Oort cloud objects would not be primordial or
directly related to the Oort cloud and thus the name inner Oort cloud
objects does not seem appropriate.

\section*{Acknowledgments}
This project used data obtained with the Dark Energy Camera (DECam),
which was constructed by the Dark Energy Survey (DES) collaborating
institutions: Argonne National Lab, University of California Santa
Cruz, University of Cambridge, Centro de Investigaciones Energeticas,
Medioambientales y Tecnologicas-Madrid, University of Chicago,
University College London, DES-Brazil consortium, University of
Edinburgh, ETH-Zurich, University of Illinois at Urbana-Champaign,
Institut de Ciencies de l'Espai, Institut de Fisica d'Altes Energies,
Lawrence Berkeley National Lab, Ludwig-Maximilians Universitat,
University of Michigan, National Optical Astronomy Observatory,
University of Nottingham, Ohio State University, University of
Pennsylvania, University of Portsmouth, SLAC National Lab, Stanford
University, University of Sussex, and Texas A\&M University. Funding
for DES, including DECam, has been provided by the U.S. Department of
Energy, National Science Foundation, Ministry of Education and Science
(Spain), Science and Technology Facilities Council (UK), Higher
Education Funding Council (England), National Center for
Supercomputing Applications, Kavli Institute for Cosmological Physics,
Financiadora de Estudos e Projetos, Fundação Carlos Chagas Filho de
Amparo a Pesquisa, Conselho Nacional de Desenvolvimento Científico e
Tecnológico and the Ministério da Ciência e Tecnologia (Brazil), the
German Research Foundation-sponsored cluster of excellence "Origin and
Structure of the Universe" and the DES collaborating institutions.
Observations were partly obtained at Cerro Tololo Inter-American
Observatory, National Optical Astronomy Observatory, which are
operated by the Association of Universities for Research in Astronomy,
under contract with the National Science Foundation.  C.T. is
supported by the Gemini observatory, which is operated by the
Association of Universities for Research in Astronomy, Inc., on behalf
of the international Gemini partnership of Argentina, Australia,
Brazil, Canada, Chile, the United Kingdom, and the United States of
America.  This research was funded by NASA Planetary Astronomy grant
NNX12AG26G and NN15AF44G. This paper includes data gathered with the
6.5 meter Magellan Telescopes located at Las Campanas Observatory,
Chile.

\newpage

%\input{SheppardTableC.tex}
%\documentstyle [aj_pt4]{article}    % Specifies the document style.

%\begin{document}

\begin{center}
\begin{deluxetable}{lccc}
\tablenum{1}
\tablewidth{6.5 in}
\tablecaption{Inner Oort Cloud Survey Observations}
\tablecolumns{4}
\tablehead{
\colhead{UT Date} & \colhead{Telescope} & \colhead{Limit} & \colhead{Area} \\ \colhead{yyyy/mm/dd} & \colhead{} & \colhead{(m$_{r}$)} & \colhead{(deg$^{2}$)}}
\startdata
2007/10/14  & Subaru    &  25.5  &  5.5   \nl
2007/10/15  & Subaru    &  24.3  &  1.0   \nl
2007/10/15  & Magellan  &  24.7  &  1.5   \nl
2007/10/16  & Magellan  &  23.3  &  1.4   \nl
2008/06/07  & Magellan  &  24.6  &  4.1   \nl
2008/06/08  & Magellan  &  24.9  &  2.2   \nl
2009/06/21  & Subaru    &  25.2  &  3.1   \nl
2009/10/14  & Subaru    &  25.7  &  8.5   \nl
2010/04/20  & Magellan  &  25.5  &  1.4   \nl
2010/04/21  & Magellan  &  25.5  &  1.6   \nl
2010/06/15  & Subaru    &  25.3  &  3.3   \nl
2010/06/16  & Subaru    &  25.5  &  1.8   \nl
2011/03/04  & Magellan  &  25.5  &  3.6   \nl
2011/03/05  & Magellan  &  25.4  &  3.4   \nl
2011/04/04  & Magellan  &  25.2  &  1.7   \nl
2011/04/04  & Magellan  &  25.5  &  1.7   \nl
2011/07/03  & Magellan  &  25.0  &  3.2   \nl
2011/09/27  & Magellan  &  25.5  &  1.8   \nl
2011/09/28  & Magellan  &  25.5  &  2.4   \nl
2012/03/16  & KPNO4m    &  24.3  &  4.7   \nl 
2012/03/17  & KPNO4m    &  24.6  &  6.4   \nl
2012/03/23  & Magellan  &  25.6  &  2.3   \nl
2012/03/24  & Magellan  &  25.6  &  2.9   \nl
2012/03/25  & Magellan  &  25.6  &  3.7   \nl
2012/07/21  & Subaru    &  25.5  &  4.5   \nl
2012/07/22  & Subaru    &  25.7  &  6.6   \nl
2012/10/12  & Magellan  &  25.3  &  1.9   \nl
2012/10/16  & Magellan  &  25.4  &  2.4   \nl
2012/11/04  & CTIO4m    &  24.0  &  10.8  \nl
2012/11/05  & CTIO4m    &  23.8  &  1.5   \nl
2012/12/11  & CTIO4m    &  24.3  &  37.8  \nl
2013/03/10  & Magellan  &  25.5  &  3.7   \nl
2013/03/11  & Magellan  &  25.3  &  3.3   \nl
2013/03/12  & Magellan  &  25.3  &  2.8   \nl
2013/03/16  & CTIO4m    &  24.6  &  64.8  \nl
2013/03/17  & CTIO4m    &  24.3  &  54.0  \nl
2013/05/08  & CTIO4m    &  24.2  &  43.2  \nl
2013/05/09  & CTIO4m    &  24.0  &  27.0  \nl
2013/08/11  & Magellan  &  25.5  &  1.2   \nl
2013/10/28  & Magellan  &  25.3  &  2.3   \nl
2013/10/29  & Magellan  &  24.8  &  0.7   \nl
2013/10/30  & Magellan  &  25.0  &  2.3   \nl
2013/11/03  & CTIO4m    &  24.0  &  32.4  \nl
2013/11/03  & CTIO4m    &  24.2  &  24.3  \nl
2013/11/04  & CTIO4m    &  24.4  &  59.4  \nl
2013/11/05  & CTIO4m    &  24.5  &  62.1  \nl
2014/03/24  & CTIO4m    &  24.8  &  64.8  \nl  
2014/03/25  & CTIO4m    &  24.7  &  54.0  \nl 
2014/03/26  & CTIO4m    &  24.5  &  59.4  \nl 
2014/03/27  & CTIO4m    &  24.5  &  37.8  \nl
2014/03/28  & CTIO4m    &  24.4  &  64.8  \nl
2014/08/29  & Subaru    &  25.7  &  6.4   \nl
2014/09/17  & CTIO4m    &  24.0  &  13.5  \nl
2014/09/18  & CTIO4m    &  24.0  &  29.7  \nl
2014/09/19  & CTIO4m    &  24.2  &  29.7  \nl
2014/09/21  & CTIO4m    &  24.4  &  59.4  \nl
2014/09/22  & CTIO4m    &  24.6  &  54.0  \nl
2015/04/13  & CTIO4m    &  24.1  &  51.3  \nl
2015/04/14  & CTIO4m    &  24.3  &  32.4  \nl
2015/11/14-15 & Magellan & 25.5  &  0.5   \nl
\enddata
\tablenotetext{}{This is an abridged version of Table 1.  To see the full Table 1, please view the paper at the Astronomical Journal where Table 1 includes all the field pointings in the survey to date.  The instruments used were SuprimeCam on Subaru, IMACS on Magellan, MOSAIC-1.1 on KPNO4m, and DECam on CTIO4m.  See the text for descriptions of the instruments and telescopes.  The limiting magnitude is the r-band magnitude where we would have found at least 50\% of the slow moving objects in the field.}
\end{deluxetable}
\end{center}

%\end{document}             % End of document.

%Need to decide about my June data in galactic plane.  Yes to some of it for sure.
%Subaru June 2010
%Subaru June 2009
%Subaru June 2008
%Magell June 2008

%Schwamb another big chunk with Subaru
%43 square degrees to 25.5 to 1200 AU reference 2009, DPS, #41, 62.06   or (Schwamb, M. and Brown, M. 2009, DPS, 41, 1124)

%Also now have Kavelaars CFHT search (Chen, Y., Kavelaars, J., Gwyn,
%S., Parker, A., Suc, V., Jordan, A., Ip, W.  2012, ACM, 6244) searched
%about 100 square degrees to 25.1 in g-band with CFHTLP data.  Found
%one object considered Sedna like, though its perihelion is not quite
%Sedna like as it comes to near 50 AU.

%Fuentes another big chunck with Magellan

%CFHT Legacy survey

%CFHT OSS survey

\newpage

%\documentstyle [aj_pt4]{article}    % Specifies the document style.

%\begin{document}

\begin{center}
\begin{deluxetable}{lcccccccccc}
%\small
\tablenum{2}
\tablewidth{6.5 in}
\tablecaption{New Extreme Distant Solar System Objects}
\tablecolumns{11}
\tablehead{
\colhead{Name} & \colhead{$q$}  &  \colhead{$a$} & \colhead{$e$}  & \colhead{$i$} & \colhead{$\Omega$} & \colhead{$\omega$} & \colhead{$b$} & \colhead{Dist}   & \colhead{Dia}  & \colhead{$m_{r}$} \\ \colhead{} & \colhead{(AU)} & \colhead{(AU)}  & \colhead{} &\colhead{(deg)} &\colhead{(deg)} & \colhead{(deg)} & \colhead{(deg)} & \colhead{(AU)}  & \colhead{(km)}  & \colhead{(mag)} }  
\startdata
\multicolumn{11}{c}{\textbf{Inner Oort Cloud}} \nl
2012 VP113     &   80.3   &   262    &    0.694  &    24.00  &     90.8      &    292.9  &  16.8     &  82.7   &  550    &   23.35  \nl
\multicolumn{11}{c}{\textbf{Extreme Detached Objects}} \nl
2014 SR349     &   48     &   288    &    0.84   &    17.98  &    34.8       &    341.3  &  17.1     &  57.2   &  200    &   24.1    \nl
2013 FT28      &   43.6   &   310    &    0.859  &    17.337 &    217.77     &    40.20  &  7.0      &  58.9   &  200    &   24.2   \nl
\multicolumn{11}{c}{\textbf{Detached/MMR+KR Objects}} \nl
2014 SS349     &   45.5   &   142    &    0.68   &    48.31  &    144.22     &    147.8  &  12.8    &  52.9   &  125    &   24.6   \nl
\multicolumn{11}{c}{\textbf{Extreme Scattered Objects}} \nl
2013 UH15      &   35.0   &   172    &    0.80   &    26.14  &    176.62     &    283.1  &  12.1     &  52.1   &  125    &   24.8   \nl
2013 FS28      &   34.5   &   196    &    0.83   &    13.01  &    204.67     &    101.5  &  2.0      &  87.9   &  400    &   24.3   \nl
\multicolumn{11}{c}{\textbf{Outer Oort Cloud}} \nl
2014 FE72      &   36.3   &   2155   &    0.98   &    20.60  &     336.77    &    134.4  &  11.0     &  59.6   &  250    &   23.7    \nl
%\hline
\enddata
\tablenotetext{}{Quantities are the perihelion ($q$), semi-major axis ($a$), eccentricity ($e$), inclination ($i$), longitude of the ascending node ($\Omega$), argument of perihelion ($\omega$), ecliptic latitude at discovery ($b$), and distance at discovery (Dist). Diameter (Dia) estimates assume a moderate albedo of 0.10.  The definitions of the various groupings are: Inner Oort Cloud ($q>50$ \& $150<a<1000$ AU), Outer Oort Cloud ($q>35 AU$ \& $a>1500$ AU), Extreme Detached Objects ($40<q<50$ \& $150<a<1000$ AU), Detached/MMR+KR Objects ($q>40$ \& $50<a<150$ AU), Extreme Scattered Objects ($30<q<40$ \& $150<a<1000$ AU).  Uncertainties are shown by the number of significant digits.}
\end{deluxetable}
\end{center}

% magnitude, size

%\end{document}             % End of document.

\newpage

%\documentstyle [aj_pt4]{article}    % Specifies the document style.

%\begin{document}

\begin{center}
\begin{deluxetable}{lccccccccc}
%\small
\tablenum{3}
\tablewidth{6.5 in}
\tablecaption{Stability of Extreme Trans-Neptunian Objects}
\tablecolumns{10}
\tablehead{
\colhead{Name} & \colhead{$q$}  & \colhead{$a$} & \colhead{$e$} &  \colhead{$i$}   & \colhead{$\Omega$}  & \colhead{$\omega$} & \colhead{$\bar{\omega}$} & \colhead{$N$} & \colhead{Stable} \\ \colhead{} & \colhead{(AU)} & \colhead{(AU)} & \colhead{} & \colhead{(deg)}  & \colhead{(deg)} & \colhead{(deg)} & \colhead{(deg)} & \colhead{(yr)} & \colhead{} }  
\startdata
\multicolumn{10}{c}{\textbf{Inner Oort Cloud}} \nl
2012 VP113 &  80.27  &   261 &  0.69 &  24.1 &  90.8  &  292.8 & 23.6  &   3 &   Yes       \nl
Sedna      &  76.04  &   499 &  0.85 &  11.9 & 144.5  &  311.5 & 96.0  &  13 &   Yes       \nl
\multicolumn{10}{c}{\textbf{Extreme Detached}} \nl    
2010 GB174 &  48.76  &   370 &  0.87 &  21.5 & 130.6  &  347.8 & 118.4 &   4 &   Yes       \nl
2014 SR349 &  47.6   &   288 &  0.84 &  18.0 &  34.8  &  341.3 & 16.1  &   1 &   Yes       \nl
2004 VN112 &  47.32  &   318 &  0.85 &  25.6 &  66.0  &  327.1 & 33.1  &   6 &   Yes       \nl
2000 CR105 &  44.29  &   226 &  0.80 &  22.7 & 128.3  &  317.2 & 85.5  &   6 &   Yes:MMR?       \nl
\bf{2013 FT28}  &  43.6   &   310 &  0.86 &  17.3 & \bf{217.8}  &   \bf{40.2} & \bf{258.0} &   3 &   Yes:Res?       \nl
\bf{2013 GP136} &  41.11  &   153 &  0.73 &  33.5 & \bf{210.7}  &   \bf{42.2} & \bf{252.9} &   3 &   Yes     \nl
\multicolumn{10}{c}{\textbf{Extreme Scattered}} \nl
\bf{2005 RH52}  &  38.98  &   151 &  0.74 &  20.5 & \bf{306.1}  &   \bf{32.3} & \bf{338.4} &   5 &   Yes:MMR?       \nl
\bf{2003 HB57}  &  38.10  &   165 &  0.77 &  15.5 & \bf{197.8}  &   \bf{10.9} & \bf{208.7} &   4 &   No       \nl
2013 RF98  &  36     &   325 &  0.88 &  29.6 & 67.6   &  316.5 & 24.1  &   0 &   No       \nl
2007 TG422 &  35.57  &   482 &  0.93 &  18.6 & 112.9  &  285.8 & 38.7  &   4 &   No       \nl
\bf{2002 GB32}  &  35.35  &   218 &  0.84 &  14.2 & \bf{177.0}  &   \bf{37.0} & \bf{214.0} &   5 &   No       \nl
2007 VJ305 &  35.18  &   188 &  0.81 &  12.0 &  24.4  &  338.3 &  2.7  &   3 &   No       \nl
2013 UH15  &  35.0   &   174 &  0.79 &  26.1 & 176.5  &  282.9 & 99.4  &   2 &   No       \nl
2010 VZ98  &  34.32  &   152 &  0.77 &   4.5 & 117.4  &  313.9 & 71.3  &   6 &   No       \nl
\bf{2001 FP185} &  34.26  &   227 &  0.85 &  30.8 & \bf{179.3}  &    \bf{7.0} & \bf{186.3} &   8 &   No       \nl
2013 FS28  &  34.1   &   199 &  0.83 &  13.1 & 204.6  &  102.0 & 306.6 &   1 &   No       \nl
2015 SO20  &  33.16  &   162 &  0.80 &  23.4 &  33.6  &  354.9 & 28.5  &   4 &   No       \nl
%\hline
\enddata
\tablenotetext{}{Orbits are from the Minor Planet Center (MPC) as of July 2016.  Bolded extreme objects are in the anti 180 deg secondary longitude of perihelion group and show opposite argument of perihelia and orbit pole angles than the main group of clustered objects. Quantities are the perihelion ($q$), semi-major axis ($a$), eccentricity ($e$), inclination ($i$), longitude of the ascending node ($\Omega$), argument of perihelion ($\omega$), longitude of perihelion ($\bar{\omega}$), and Number of years observed.  An object is considered stable if it moved less than a few AU in semi-major axis during our 1 billion year numerical integrations.  2000 CR105 and 2005 RH52 showed stable semi-major axes, but the stepped motion of $a$ over time suggests they could be in a mean motion resonance with Neptune.  2013 FT28 also had some jumpiness in its semi-major axis, usually about 10 AU for the nominal and 1 sigma clones orbits, suggesting some significant interactions with Neptune, see text.  All unstable objects had well over 100 AU changes in their semi-major axes within 1 billion years, with most lost within about 1 billion years including 2010 VZ98 and 2007 TG422.}
\end{deluxetable}
\end{center}

%\end{document}             % End of document.

\newpage

%\documentstyle [aj_pt4]{article}    % Specifies the document style.

%\begin{document}

\begin{center}
\begin{deluxetable}{lcccccccccc}
%\small
\tablenum{4}
\tablewidth{6.5 in}
\tablecaption{Other New Interesting Distant Solar System Objects}
\tablecolumns{11}
\tablehead{
\colhead{Name} & \colhead{$q$}  &  \colhead{$a$} & \colhead{$e$}  & \colhead{$i$} & \colhead{$\Omega$} & \colhead{$\omega$} &  \colhead{Dist}   & \colhead{Dia}  & \colhead{$m_{r}$} & \colhead{$N:N$} \\ \colhead{} & \colhead{(AU)} & \colhead{(AU)}  & \colhead{} &\colhead{(deg)} &\colhead{(deg)} & \colhead{(deg)} & \colhead{(AU)}  & \colhead{(km)} & \colhead{(mag)} & \colhead{} }  
\startdata
%\multicolumn{11}{c}{\textbf{Possible Neptune Resonance}} \nl
2014 SW349        &   39     &   96     &    0.59   &    15.2   &    19.8     &    251.7  & 50.1    &  100     &   24.5  &      17:3       \nl
2014 FJ72         &   38     &   94     &    0.59   &    15.4   &    302.7    &    132.7  & 69.3    &  300     &   23.9  &      11:2/17:3   \nl
2014 FL72         &   38     &   103    &    0.63   &    29.17  &    201.14   &    258.99 & 62.4    &  175     &   24.6  &      6:1/13:2       \nl
2014 SZ349        &    37    &   79     &    0.54   &    35.8   &    152.9     &   135.6  & 50.9    &  125     &   24.7  &      13:3       \nl
2014 SD350        &    34    &   79     &    0.57   &    32.8   &    356.6     &   46.6   & 48.2    &  150     &   24.0  &      13:3       \nl
%\hline
\enddata
\tablenotetext{}{Quantities are the perihelion ($q$), semi-major axis ($a$), eccentricity ($e$), inclination ($i$), longitude of the ascending node ($\Omega$), argument of perihelion ($\omega$), distance at discovery (Dist), and location near Neptune resonance ($N:N$). Diameter (Dia) estimates assume a moderate albedo of 0.10. The uncertainty of the orbital elements for each object are shown by the number of significant digits.}
\end{deluxetable}
\end{center}

%\end{document}             % End of document.

\newpage

\begin{table}
  \tablenum{5}
  \footnotesize
  \centering
  \begin{tabular}{|l|r|l|} \hline 
    \multicolumn{3}{|l|}{Common Parameters} \\
    \hline
    Parameter     & Value   & Description     \\
    \hline
    $\rho$        & 1000 ${\mbox{kg }}{\mbox{m}}^{-3}$ & Density \\
    $p_r        $ & 0.10                     & Albedo in $r$ filter\\
    $r_{\sf min}$ & 75 km                    & Minimum radius \\
    $r_{\sf max}$ & 1200 km                  & Maximum radius \\
    $a_{\sf max}$ & 800 AU                   & Maximum semi-major axis \\
    $e_{\sf min}$ & 0.625                    & Minimum eccentricity (uniform distribution) \\
    $q'         $ & 4 or 5                   & Size distribution power law exponent \\
    $a          $ & 1                        & Semi-major axis distribution power law exponent \\
    $\sigma_i   $ & 6.9$^\circ$              & Sigma for Gaussian inclination distribution \\
    $\mu        $ & 19.1$^\circ$             & Mean for Gaussian inclination distribution \\
    $\omega_{\sf min}$ & -80 deg             & Minimum argument of perihelion \\
    $\omega_{\sf max}$ & 45 deg              & Maximum argument of perihelion \\
    \hline
    \hline
    \multicolumn{3}{|l|}{Inner Oort Cloud Parameters} \\
    \hline
    $q_{\sf min}$ & 75 AU                   & Minimum perihelion \\
    $a_{\sf min}$ & 200 AU                   & Minimum semi-major axis \\
    $\bar{\omega}_{\sf min}$ & 0 deg         & Minimum longitude of perihelion \\
    $\bar{\omega}_{\sf max}$ & 130 deg       & Maximum longitude of perihelion \\
    $N_{\sf obs}$ & 1  & Observed objects: 2012 $\rm VP_{113}$ \\
    $N$ ($q' = 4$) & 9.4 $\times$ 10$\sf ^3$  & Best-fit number of objects $r > 100$ km \\
    $N$ ($q' = 5$) & 2.1 $\times$ 10$\sf ^4$  & Best-fit number of objects $r > 100$ km \\
    \hline
    \hline
    \multicolumn{3}{|l|}{Extreme Detached Objects and Extreme Scattered Objects ($\bar{\omega}$ Constrained)} \\
    \hline
    $q_{\sf min}$ & 35 AU             & Minimum perihelion \\
    $q_{\sf max}$ & 50 AU             & Maximum perihelion \\
    $a_{\sf min}$ & 150 AU                   & Minimum semi-major axis \\
    $\bar{\omega}_{\sf min}$ & 0 deg         & Minimum longitude of perihelion \\
    $\bar{\omega}_{\sf max}$ & 130 deg       & Maximum longitude of perihelion \\
    $N_{\sf obs}$ & 2  & Observed objects: c198c45, M1670c5 \\
    $N$ ($q' = 4$) & 1900  & Best-fit number of objects $r > 100$ km \\
    $N$ ($q' = 5$) & 1800  & Best-fit number of objects $r > 100$ km \\
    \hline
    \hline
    \multicolumn{3}{|l|}{Extreme Detached Objects and Extreme Scattered Objects ($\bar{\omega}$ Unconstrained)} \\
    \hline
    $q_{\sf min}$ & 35 AU             & Minimum perihelion \\
    $q_{\sf max}$ & 50 AU             & Maximum perihelion \\
    $a_{\sf min}$ & 150 AU                   & Minimum semi-major axis \\
    $\bar{\omega}_{\sf min}$ & 0 deg         & Minimum longitude of perihelion \\
    $\bar{\omega}_{\sf max}$ & 360 deg       & Maximum longitude of perihelion \\
    $N_{\sf obs}$ & 3  & Observed objects: c198c45, D292c54, M1670c5 \\
    $N$ ($q' = 4$) & 2900  & Best-fit number of objects $r > 100$ km \\
    $N$ ($q' = 5$) & 2700  & Best-fit number of objects $r > 100$ km \\
\hline
    \end{tabular}
  \caption{Population parameters used for the observational bias
    simulations. The actual number of simulated objects was at least a
    factor 10 greater than the best fit number of detected simulated
    objects in all cases. We report the number of objects for radii
    greater than 100 km for the various simulation
    parameters. Radially asymmetric and symmetric populations (uniform
    in $\omega$ and $\bar{\omega}$) were considered for all
    populations.}
\end{table}

\newpage

\newpage

\newpage

\newpage

\newpage

\newpage

\newpage

\newpage

\newpage

\newpage

\newpage 

\newpage

\newpage

\clearpage

\newpage

%\newpage 
%
%\begin{figure}
%\epsscale{0.4}
%\centerline{\includegraphics[angle=90,totalheight=0.6\textheight]{ESDOplan2016.ps}}
%\caption{Same as Figure~\ref{fig:ETNOplan2016}, except including the
%  lower perihelion ($35<q<40$ AU) extreme scattered disk objects in
%  green.  It is interesting that the extreme scattered objects with
%  similar longitudes of perihelion as the IOCs and EDTNOs have
%  increasing aphelion that correlate with a slightly }
%\label{fig:ESDOplan2016} 
%\end{figure}

\newpage

\newpage

\newpage

\newpage

\end{document}